\documentclass[
reprint, 
 amsmath,amssymb,
 aps,
prb,
]{revtex4-2}

\usepackage{dcolumn}
\usepackage{graphicx}
\usepackage[bookmarks=false,colorlinks=true,urlcolor=blue,citecolor=blue,linkcolor=blue,breaklinks=true]{hyperref}
\usepackage{color}
\usepackage{ascmac,amsmath,amsfonts,amsthm,amssymb}
\usepackage{url}
\usepackage{siunitx}

\usepackage{bm}

\usepackage{txfonts}

\newcommand{\n}{\nonumber \\}

\def\ket#1{|#1\rangle }
\def\bra#1{\langle #1 |}
\def\braket#1{\langle #1 \rangle}
\def\n{\nonumber \\ }

\begin{document}

\title{
Formulation of the orbital magnetic moment in multiorbital tight-binding models: \\
Application to the inverse Faraday effect}

\author{Kosuke Tazuke}
\affiliation{%
 Department of Applied Physics, The University of Tokyo, Hongo, Tokyo, 113-8656, Japan
}%

\author{Takahiro Morimoto}
\affiliation{%
 Department of Applied Physics, The University of Tokyo, Hongo, Tokyo, 113-8656, Japan
}%

\author{Sota Kitamura}
\affiliation{%
 Department of Applied Physics, The University of Tokyo, Hongo, Tokyo, 113-8656, Japan
}%

\date{\today}

\begin{abstract}
    We establish a theoretical formulation of the orbital magnetic moment in multiorbital tight-binding models, focusing on the role of the electric dipole. We demonstrate that the total magnetic moment can be decomposed into several contributions in multiorbital tight-binding models generally. In particular, we reveal that the electric dipole moment of Wannier orbitals also contributes to the orbital magnetic moment, which is not included in the conventional expression for the orbital magnetic moment in lattice systems. The derived formulation for the magnetic moment is applied to the inverse Faraday effect (IFE), a phenomenon where circularly-polarized light induces a magnetic moment. To account for all possible contributions, we adopt an $s$-$p$ tight-binding system as a minimal model for studying the IFE. Using an analytical approach based on the Schrieffer-Wolff transformation, we clarify the physical origins of these contributions. Additionally, we quantitatively evaluate each contribution on an equal footing through a numerical approach based on the Floquet formalism. Our results reveal that the orbital magnetic moment exhibits a significantly larger response compared to the spin magnetic moment, with all contributions to the orbital magnetic moment being comparable in magnitude. These findings highlight the essential role of orbital degrees of freedom in the IFE.
\end{abstract}

\maketitle

\section{Introduction}\label{sec : Introduction}

\begin{figure}[t]
    \centering
    \includegraphics[width = \linewidth]{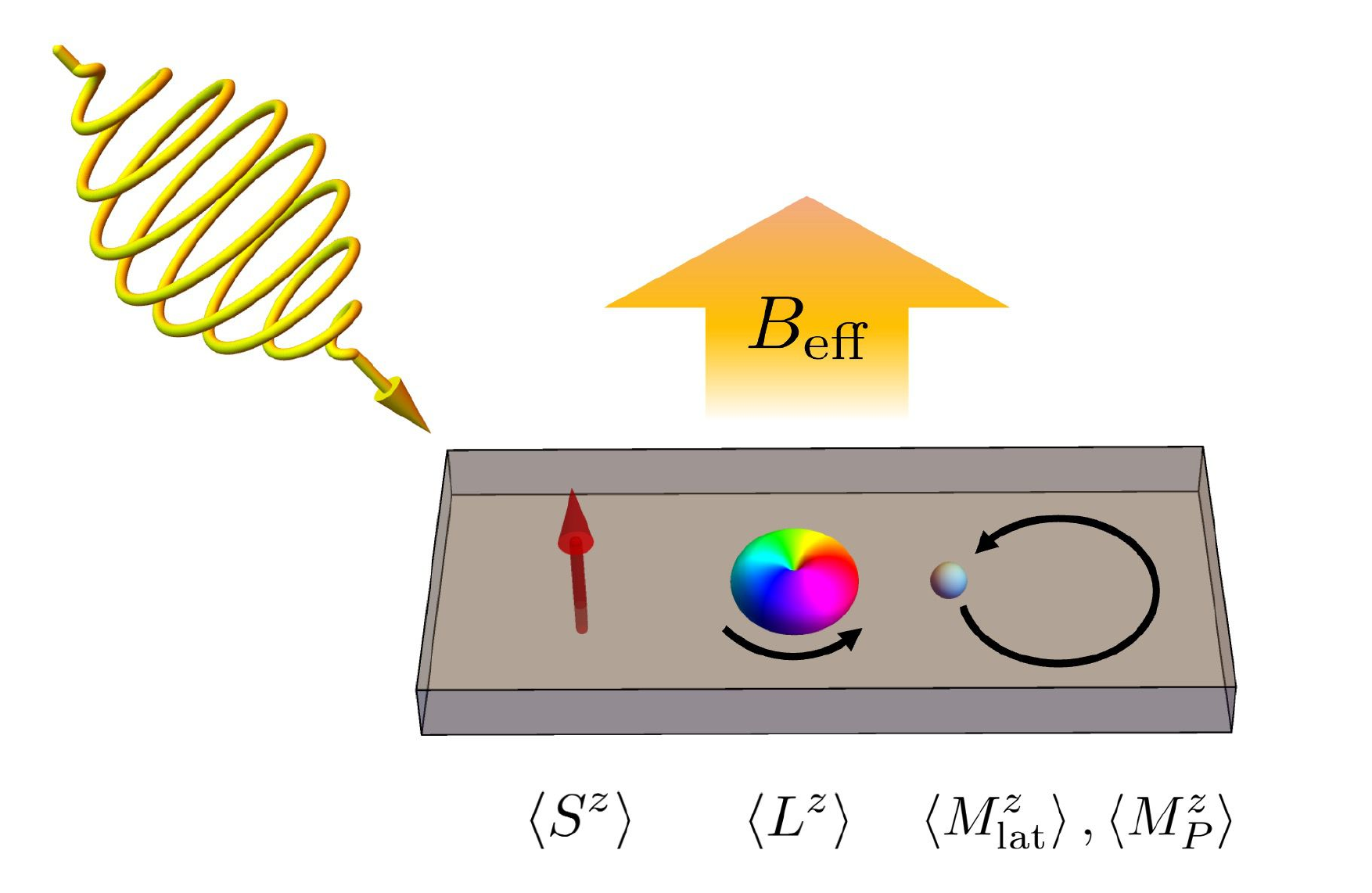}
    \caption{ 
    Schematic picture of the inverse Faraday effect. 
    When circularly-polarized light is irradiated into the material, an effective magnetic field is generated, inducing spin angular momentum $\braket{S^z}$, orbital angular momentum $\braket{L^z}$ and kinetic orbital magnetic moment $\braket{M^z_{\text{lat}}}$ and $\braket{M^z_{P}}$.}
    \label{fig: IFE}
\end{figure}

Probing and controlling magnetism in solids via light has long been one of the central issues in condensed matter physics \cite{Kirilyuk2010}.
The inverse Faraday effect (IFE) is one of the magneto-optical effects where static magnetization is induced in materials by irradiation with circularly-polarized light (Fig.~\ref{fig: IFE}).
It is a kind of nonlinear optical response as the magnetization $\bm{M}_{\text{tot}}$ is proportional to the square of the electric field $\bm{E}$, which can be described by $\bm{M}_{\text{tot}}(0) \propto \bm{E}(\omega) \times \bm{E}(-\omega)$.
With the help of recently developed laser technology, this effect has been observed in various materials in the experiments such as pump-probe measurements \cite{Kimel2005,Hansteen2006,Satoh2010,Makino2012,Kozhaev2018,Amano2022,Ortiz2023} and spin current detection techniques \cite{Kawaguchi2020}. The IFE has attracted keen attention for its potential applications in optospintronics, as it allows for the ultrafast control of magnetization in materials. Consequently, it has been extensively studied from both theoretical and experimental perspectives.

The physical origin and formulation of the IFE has been a subject of theoretical discussion for many years \cite{Pitaevskii1961,Pershan1966,Riccardo2006,Battiato2014,Watanabe2024}. Theoretical studies were performed in various systems, including normal metals \cite{Riccardo2006,Tatara2011,Karakhanyan2022,Mishra2023,Sharma2024}, topological materials \cite{Misawa2011,Tokman2020,Gao2020,Liang2021,Qu2022,Zhang2023,Cao2024}, the electron systems with Rashba spin-orbit coupling \cite{Tanaka2020,Tanaka2024}, superconductors \cite{Mironov2021,Majedi2021,Dzero2024}, Mott insulators \cite{Banerjee2022}, and quantum spin systems \cite{Takayoshi2014,Takayoshi2014-2,Dannegger2021}, based on various approaches including classical analysis, perturbation theory, Green function methods, and Floquet theory, often combined with time evolution simulations. Additionally, first-principles calculations have also been employed in these investigations \cite{Berritta2016,Dannegger2021,Xu2021,Mishra2023,Adamantopoulos2024,Mishra2024}.

The magnetic moment in quantum systems consists of the spin and orbital contributions, which are usually observed without distinction in experiments.
This implies that a systematic comparison of both contributions based on microscopic theory is necessary to understand the underlying mechanism, whereas most previous theoretical studies have focused on a particular contribution. 
On the other hand, for the orbital contribution in the (multiorbital) tight-binding model, the definition of the orbital magnetic moment has been unclear so far; 
While the magnetic moment arising from the kinetic motion of electrons is known to be described by the so-called modern theory of orbital magnetization \cite{Xiao2005,Thonhauser2005,Ceresoli2006,Xiao2006,Shi2007,Xiao2010,Resta2010,Topp2022}, this evidently fails to capture the contribution from the orbital angular momentum expressed by the internal orbital degrees of freedom. 

In this study, we first clarify that the orbital magnetic moment can be decomposed into three contributions. Specifically, with the spin magnetic moment, the total magnetic moment in the tight-binding model should be defined as
\begin{equation}
\bm{M}_{\text{tot}}=-\frac{\mu_B}{\hbar}\langle 2\bm{S}+\bm{L}\rangle+\bm{M}_{\text{lat}}+\bm{M}_{P},\label{eq:magnetization}
\end{equation}
where $\bm{S}$ and $\bm{L}$ are the spin and orbital angular momentum of each Wannier orbital (internal orbital degree of freedom), while $\bm{M}_{\text{lat}}$ describes the orbital magnetization due to the kinetic motion, which is described by the modern theory and coupled to the magnetic field introduced via the Peierls phase. We find that an overlooked contribution, $\bm{M}_{P}$, also arises due to the electric dipole moment of the Wannier orbital. 
Then, we investigate all the contributions to the magnetic moment on an equal footing, and discuss their origins and differences using an $s$-$p$ tight-binding model as a minimal model for the IFE. This allows us to elucidate the effects of multiorbital characteristics on the IFE, which could not be addressed in previous model calculations lacking an appropriate treatment of the orbital degrees of freedom. We employ the Floquet formalism to describe the IFE, with which the spin magnetic moment arises from the emergent Zeeman term in the Floquet Hamiltonian, while the orbital magnetic moment can be described by an effective operator. Combined with the numerical calculation, we reveal that the orbital magnetic moment exhibits a stronger response compared to the spin magnetic moment, and all three contributions to the orbital magnetic moment are found to be comparable in magnitude.

The rest of the paper is organized as follows. 
In Sec.~\ref{sec : OM}, we begin by discussing that the electric polarization and the magnetic moment in multiorbital tight-binding models can be decomposed into those coupled to the Peierls phase and the correction term representing those of the Wannier orbital. Specifically, we show that the total magnetic moment can be decomposed into four contributions.
In Sec.~\ref{sec : Setup}, we introduce the $s$-$p$ tight-binding model under circularly-polarized light. We briefly review the Floquet formalism and the method to calculate the expectation values for the steady state coupled to the fermionic reservoir.
In Sec.~\ref{sec : Analytical Result}, we show analytical results derived from the time-dependent Schrieffer-Wolff transformation to discuss the origin of magnetic moments. 
In Sec.~\ref{sec : Numerical Result}, we present numerical results of the Floquet band structure and the frequency dependence of the IFE.
We also discuss the interpretation of the results, their correspondence with analytical results, and the differences between the four types of magnetization.
Finally, we provide a brief discussion in Sec.~\ref{sec : Discussion}.



\section{Orbital magnetization in multiorbital systems}\label{sec : OM}

In this section, we revisit the formulation of the orbital magnetization as well as the electric polarization. We show that the correction terms due to the Wannier orbital are necessary in multiorbital tight-binding systems, which have not been fully recognized so far. Let us consider a generic
lattice Hamiltonian of the form 
\begin{equation}
H=\sum_{ijll^\prime}c_{il}^{\dagger}H_{ll^\prime,ij}c_{jl^\prime},
\end{equation}
where $c_{il}$ denotes the field operator of the electron at $i$th
unit cell. Here the internal degrees of freedom including orbital and spin
are indicated altogether by the label $l$ for simplicity. We assume
that the lattice Hamiltonian is translationally invariant, so
that it is Fourier-transformed into $H=\sum_{\bm{k}ll^\prime}c_{\bm{k}l}^{\dagger}H_{ll^\prime}(\bm{k})c_{\bm{k}l^\prime}$ with $c_{\bm{k}l^\prime}=N^{-1/2}\sum_{j}c_{jl^\prime}e^{-i\bm{k}\cdot\bm{R}_{j}}$
and $H_{ll^\prime}(\bm{k})=N^{-1}\sum_{ij}H_{ll^\prime,ij}e^{-i\bm{k}\cdot(\bm{R}_{i}-\bm{R}_{j})}$.
Here $N$ is the number of unit cells and $\bm{R}_{i}$ is the
center position of $i$th unit cell.

\subsection{Wannier correction for the electric dipole moment}
Let us begin with the expression for the electric dipole.
According to the modern theory of polarization \cite{King-Smith1993,Resta1994}, the matrix element of the electric dipole
in a crystalline system $-e\bm{a}$ is given by 
\begin{align}
-e\bm{a}_{\alpha\beta}(\bm{k}) & =-e\frac{1}{N}\int d^{3}\bm{r}u_{\alpha\bm{k}}^{\ast}(\bm{r})i\frac{\partial u_{\beta\bm{k}}(\bm{r})}{\partial\bm{k}},\label{eq:dipole-continuous}
\end{align}
where $u_{\alpha\bm{k}}(\bm{r})$ denotes the periodic part of the
Bloch wave function with eigenenergy $E_{\alpha\bm{k}}$. Although this quantity itself is not gauge invariant, the electric polarization of band $\alpha$ is obtained as $-e\int d^3\bm{k} \bm{a}_{\alpha\alpha}$. In the tight-binding model coupled to the electric
field via the Peierls phase, the counterpart of the above formula,
denoted by $-e\bm{a}_{\text{lat}}$ here, is obtained as
\begin{equation}
-e[\bm{a}_{\text{lat}}(\bm{k})]_{\alpha\beta}=-e\sum_{l}\phi_{\alpha\bm{k}l}^{\ast}i\frac{\partial\phi_{\beta\bm{k}l}}{\partial\bm{k}},
\end{equation}
where $\phi_{\alpha\bm{k}l}$ is the one-particle eigenstate of the lattice Hamiltonian with
eigenenergy $E_{\alpha}(\bm{k})$, i.e., $\sum_{l^\prime}H_{ll^\prime}(\bm{k})\phi_{\alpha\bm{k}l^\prime}=E_{\alpha\bm{k}}\phi_{\alpha\bm{k}l}$.
While these two expressions are often supposed to be
equivalent, as we see below, we need to take account of their difference as the Wannier correction in multiorbital
problems. 

In terms of the Bloch electrons, $c_{il}$
corresponds to the field operator for the $l$th maximally-localized
Wannier orbital at $i$th site, $w_{l}(\bm{r}-\bm{R}_{i})$. This implies
that the eigenstate $\phi_{\alpha\bm{k}l}$ in the lattice system
can be mapped to the Bloch wave function $u_{\alpha\bm{k}}(\bm{r})$
in the continuous space as
\begin{equation}
u_{\alpha\bm{k}}(\bm{r})e^{i\bm{k}\cdot\bm{r}}=\sum_{l}\phi_{\alpha\bm{k}l}\sum_{i}w_{l}(\bm{r}-\bm{R}_{i})e^{i\bm{k}\cdot\bm{R}_{i}},\label{eq:bloch-to-lattice}
\end{equation}
see details for Appendix~\ref{sec:appendix-wannier}. 
Then, substituting this expression into Eq.~(\ref{eq:dipole-continuous}) yields
\begin{align}
-e\bm{a}_{\alpha\beta}(\bm{k}) & =-e[\bm{a}_{\text{lat}}(\bm{k})]_{\alpha\beta}+\sum_{ll^\prime}\phi_{\alpha\bm{k}l}^{\ast}\bm{P}_{ll^\prime}(\bm{k})\phi_{\beta\bm{k}l^\prime},\label{eq:dipole-operator}
\end{align}
where 
\begin{equation}
\bm{P}_{ll^\prime}(\bm{k})=-e\sum_{i}\int d^{3}\bm{r}w_{l}^{\ast}(\bm{r})\bm{r}w_{l^\prime}(\bm{r}+\bm{R}_{i})e^{-i\bm{k}\cdot\bm{R}_{i}}
\label{eq : Wannier polarization}
\end{equation}
is interpreted as the electric dipole moment of the Wannier orbital. 

We note that the existence of the Wannier correction to the electric dipole implies that the coupling of the electric field to the lattice Hamiltonian should also have a correction. As explicitly shown in Appendix~\ref{sec:appendix-dipole}, the dipole interaction
\begin{equation}
H_{\text{dipole}}=-\sum_{\bm{k}ll^\prime}c^\dagger_{\bm{k}l}(\bm{E}(t)\cdot\bm{P}_{ll^\prime}(\bm{k}))c_{\bm{k}l^\prime}\label{eq:dipole-interaction}
\end{equation}
appears as a correction term when projecting the Bloch Hamiltonian in the presence of the electric field onto the tight-binding subspace.

\subsection{Wannier correction for the orbital magnetization}
In a similar way, we reveal that the orbital magnetization also has the Wannier
correction. As we detailed in Appendix~\ref{sec:appendix-OM}, the orbital magnetic moment in terms of the Bloch electrons $\bm{M}_{\text{orb}}$ in equilibrium can be decomposed into three contributions as
\begin{align}
\bm{M}_{\text{orb}} & =\bm{M}_{\text{lat}}+\bm{M}_{L}+\bm{M}_{P},\label{eq:orbitalmag-decomp}
\end{align}
where $\bm{M}_{\text{lat}}$ is the kinetic contribution in terms of the lattice Hamiltonian,
\begin{align}
\bm{M}_{\text{lat}} & =\frac{ie}{2\hbar}\sum_{\bm{k}ll^\prime\alpha}\frac{\partial\phi_{\alpha\bm{k}l}^{\ast}}{\partial\bm{k}}\times(\delta_{ll^\prime}E_{\alpha\bm{k}}-H_{ll^\prime})\frac{\partial \phi_{\alpha\bm{k}l^\prime}}{\partial\bm{k}}f_\alpha\nonumber \\
 & -\frac{ek_BT}{\hbar}\sum_{\bm{k}\alpha}\frac{\partial}{\partial\bm{k}}\times[\bm{a}_{\text{lat}}]_{\alpha\alpha}\ln(1-f_\alpha),
 \label{eq : lattice-contribution}
\end{align}
which is exact for the magnetic field introduced via the Peierls phase, while the remaining two terms represent the Wannier correction. Here $f_\alpha=f(E_{\alpha\bm{k}})$ is the Fermi distribution function. In multiorbital
systems, the Wannier orbital itself can possess the magnetic
moment as the orbital angular momentum, which is described by the second term
\begin{align}
\bm{M}_{L} & =-\frac{\mu_{B}}{\hbar}\sum_{\bm{k}ll^\prime}\langle c_{\bm{k}l}^{\dagger}\bm{L}_{ll^\prime}c_{\bm{k}l^\prime}\rangle
\end{align}
with $\mu_B$ being the Bohr magneton, where 
\begin{align}
\frac{\bm{L}_{ll^\prime}}{\hbar} & =-i\sum_{i}\int d^3\bm{r}w_{l}^{\ast}(\bm{r})\left(\bm{r}+\frac{\bm{R}_{i}}{2}\right)\times\bm{\nabla}w_{l^\prime}(\bm{r}+\bm{R}_{i})
e^{-i\bm{k}\cdot\bm{R}_{i}}\label{eq:angular-momentum-wannier}
\end{align}
represents the orbital angular momentum of the Wannier orbital. The expectation value here can be explicitly computed as $\langle c_{\bm{k}l}^\dagger c_{\bm{k}l^\prime} \rangle = \sum_{\alpha}\phi_{\alpha\bm{k}l}^{\ast}\phi_{\alpha\bm{k}l^\prime}f_\alpha$.
We have another
correction term due to the dipole moment of the Wannier orbital, expressed as
\begin{align}
\bm{M}_{P} & =\frac{1}{2}\text{Re}\sum_{\bm{k}ll^\prime l^{\prime\prime}}\langle c_{\bm{k}l}^{\dagger}(\bm{P}_{ll^{\prime\prime}}\times\bm{v}_{l^{\prime\prime}l^\prime})c_{\bm{k}l^\prime}\rangle,
\end{align}
with $\bm{v}_{ll^\prime}=\hbar^{-1}\partial_{\bm{k}}H_{ll^\prime}$ being the velocity
operator for the lattice Hamiltonian. 
The physical meaning of $\bm{M}_{P}$ is as follows.
The classical expression for the orbital magnetic moment is given by $\bm{M}=-e(\bm{r}\times\bm{v})/2$, which is evaluated for the electron wave packet~\cite{Xiao2005} as $\bm{M}_{\text{lat}}$, when the system has no orbital degree of freedom. 
When the wave packet consists of the orbital with nonzero dipole moment $\bm{P}$ (e.g., $s+p$ orbital), the center-of-mass position of the wave packet is shifted by $\delta\bm{r}=-\bm{P}/e$, which modulates the magnetic moment by $\delta\bm{M}=(\bm{P}\times\bm{v})/2$. This is nothing but the correction term $\bm{M}_P$ derived above.

The geometric contribution $\bm{M}_{\text{lat}}$ plays a dominant role in single-orbital models, while $\bm{M}_L=-\mu_B\langle\bm{L}\rangle/\hbar$ survives even at the atomic limit and is substantial in localized systems.
However, as we demonstrate in this paper, both contributions, as well as the overlooked contribution $\bm{M}_{P}$, can be significant in typical situations, and an equal-footing treatment is essential.

Before closing the section, let us comment on the present expression with the formulation given in Ref.~\cite{Lopez2012}, where the methodology to reproduce the first-principles magnetic moment in the Wannier interpolation scheme (i.e., in the downfolded tight-binding model) is provided. 
Indeed, our formulation is equivalent to theirs for the total magnetic moment, while we adopt a different decomposition scheme.
In their approach, the magnetic moment is decomposed into local and itinerant circulations~\cite{Ceresoli2006}, $\bm{M}_{\text{LC}}$ and $\bm{M}_{\text{IC}}$, which are known to be invariant against the arbitrariness of the Wannier orbitals, respectively. The correction terms expressed within the Wannier space are provided for the two contributions, which are suitable for the quantitative computation.

On the other hand, our formalism adopts the decomposition based on the conventional expressions, and written in terms of the physical quantities, atomic orbital angular momentum $\bm{L}$ and dipole moment $\bm{P}$. Our formalism is advantageous in that, as in the Slater-Koster parameterization for the hopping amplitudes, we can determine the form of $\bm{L}$ and $\bm{P}$ as matrices with a few free parameters, without specifying material. 
This can be done by approximating the angular dependence of the Wannier orbital by one spherical harmonic (that of the atomic orbital). 
In particular, when only the onsite matrix elements [$\bm{R}_i=0$ term in Eqs.~(\ref{eq : Wannier polarization}) and (\ref{eq:angular-momentum-wannier})] are considered, the matrix form of $\bm{L}$ and $\bm{P}$ are considerably simplified. For instance, in the subspace spanned by the $s$ and $p$ orbitals, $\bm{L}$ is given by the standard matrix representation as
\begin{equation}
\frac{L^{x}}{\hbar}=\begin{pmatrix}0 & 0 & 0 & 0\\
0 & 0 & 0 & 0\\
0 & 0 & 0 & -i\\
0 & 0 & i & 0
\end{pmatrix},\frac{L^{y}}{\hbar}=\begin{pmatrix}0 & 0 & 0 & 0\\
0 & 0 & 0 & i\\
0 & 0 & 0 & 0\\
0 & -i & 0 & 0
\end{pmatrix},\frac{L^{z}}{\hbar}=\begin{pmatrix}0 & 0 & 0 & 0\\
0 & 0 & -i & 0\\
0 & i & 0 & 0\\
0 & 0 & 0 & 0
\end{pmatrix},\label{eq:onsite-l}
\end{equation}
while $\bm{P}$ follows the Laporte rule (i.e., nonzero when parity changes) and given by
\begin{equation}
P^{x}=\begin{pmatrix}0 & P & 0 & 0\\
P & 0 & 0 & 0\\
0 & 0 & 0 & 0\\
0 & 0 & 0 & 0
\end{pmatrix},P^{y}=\begin{pmatrix}0 & 0 & P & 0\\
0 & 0 & 0 & 0\\
P & 0 & 0 & 0\\
0 & 0 & 0 & 0
\end{pmatrix},P^{z}=\begin{pmatrix}0 & 0 & 0 & P\\
0 & 0 & 0 & 0\\
0 & 0 & 0 & 0\\
P & 0 & 0 & 0
\end{pmatrix},\label{eq:onsite-p}
\end{equation}
where the basis vectors are $s,p_x,p_y,p_z$ in order.
This implies that $\bm{M}_P$ can have a substantial contribution when the orbitals with different parity are present in the model.


\section{Model and methods}\label{sec : Setup}

In Sec.~\ref{sec : OM}, we demonstrated that the total magnetic moment in multiorbital tight-binding models can be decomposed into four distinct contributions. In this section, we introduce a minimal model to study the IFE in multiorbital systems, allowing for the evaluation of all four contributions on an equal footing.

\subsection{$s$-$p$ tight-binding model}

\begin{figure}[t]
    \centering
    \includegraphics[width = \linewidth]{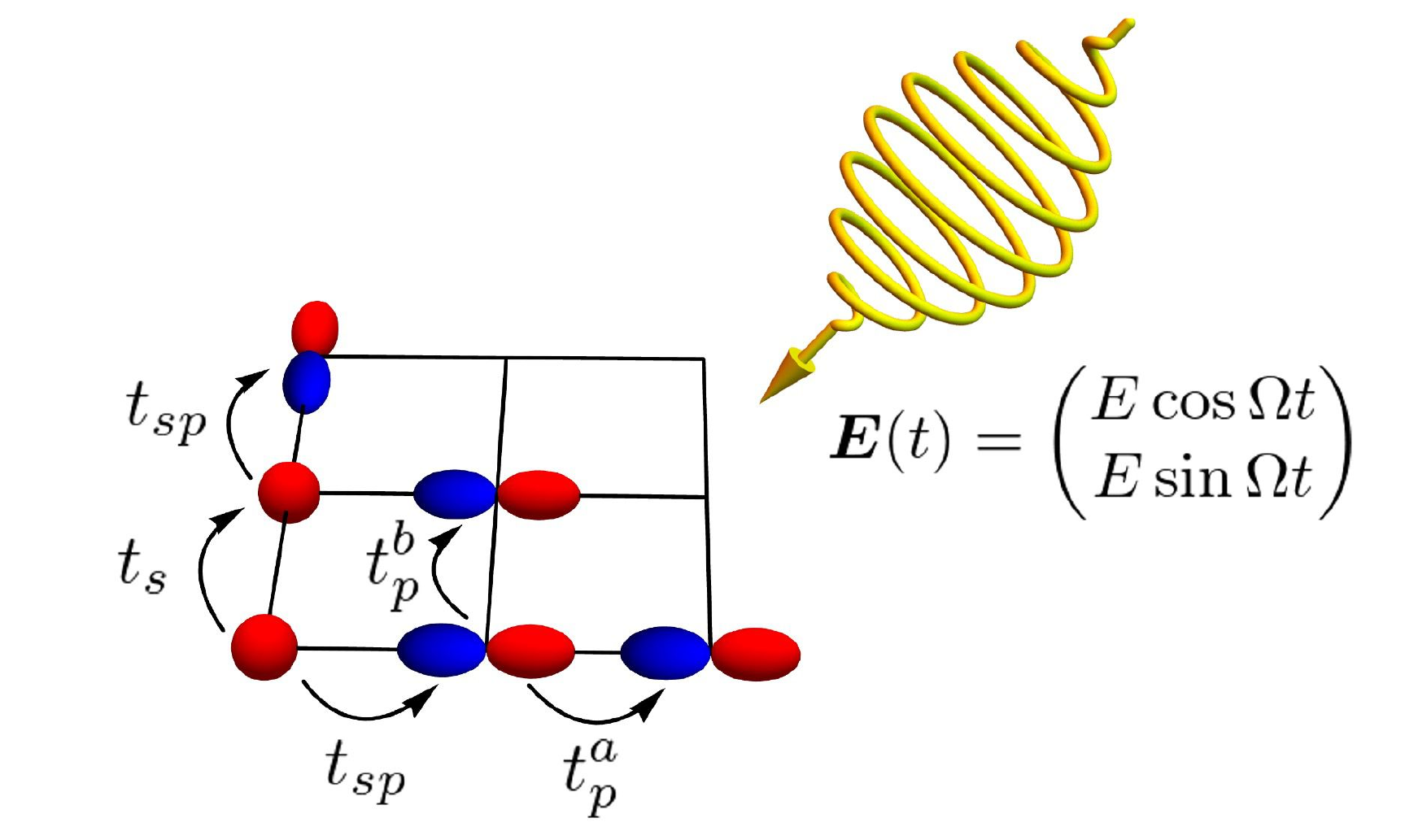}
    \caption{Schematic picture of the model. The system is defined on a square lattice with $s$, $p_x$, and $p_y$ orbitals. The system is irradiated with circularly-polarized light.}
    \label{fig: Model}
\end{figure}

To capture all four contributions in Eq.~(\ref{eq:magnetization}) in a minimal manner, we concentrate on the tight-binding model defined on the square lattice with $s$, $p_x$ and $p_y$ orbitals in this study (Fig.~\ref{fig: Model}).
Here, the multiorbital nature is necessary to have nonzero contributions from the orbital angular momentum $\bm{L}$ [Eq.~(\ref{eq:angular-momentum-wannier})] and the electric dipole moment $\bm{P}$ [Eq.~(\ref{eq : Wannier polarization})]. As we have seen in Sec.~\ref{sec : OM}, two $p$ orbitals (a pair of $l\ge1$ states in general) are necessary to have onsite matrix elements of $\bm{L}$, while nonzero $\bm{P}$ may arise when orbitals with different parity ($s$ and $p$ here) are considered [See Eqs.~(\ref{eq:onsite-l}), (\ref{eq:onsite-p})]. Thus, the $s, p_x$ and $p_y$ orbitals are a minimal set having both contributions.

The tight-binding Hamiltonian in the presence of the applied electric field is given by
\begin{align}
    H &= H_{s} + H_{p} + H_{sp} + H_{\text{SOC}} + H_{\text{dipole}},
\end{align}
where $H_s$ and $H_p$ denote the intraorbital hopping terms
\begin{align}
    H_{s} &= -t_s \sum_{\bm{r},\sigma,i}
    \left(
    e^{-i\bm{A}(t)\cdot \bm{e}_i}  
        c^{\dagger}_{\bm{r}+\bm{e}_{i},\sigma,s}c_{\bm{r},\sigma,s}
        + \text{h.c.}
    \right),\\
    H_{p} &= \Delta\sum_{\bm{r},\sigma,i}c^{\dagger}_{\bm{r},\sigma,p_i}c_{\bm{r},\sigma,p_i} \n
    &- t_{p}^{a}\sum_{\bm{r},\sigma,i}
    \left(
        e^{-i\bm{A}(t)\cdot \bm{e}_i}c^{\dagger}_{\bm{r}+\bm{e}_i,\sigma,p_i}c_{\bm{r},\sigma,p_i}
        + \text{h.c.}
    \right)   \n
    &
    - t_{p}^{b}\sum_{\bm{r},\sigma,i\neq j}
    \left(
        e^{-i\bm{A}(t)\cdot \bm{e}_i}c^{\dagger}_{\bm{r}+\bm{e}_i,\sigma,p_j}c_{\bm{r},\sigma,p_j}
        +
        \text{h.c.}
    \right),
\end{align}
while $H_{sp}$ describes the interorbital hopping
\begin{align}
    H_{sp} &= -t_{sp}\sum_{\bm{r},\sigma,i}
    (
    e^{-i\bm{A}(t)\cdot \bm{e}_i}c^{\dagger}_{\bm{r}+\bm{e}_i,\sigma,p_i}c_{\bm{r},\sigma,s}  
    +
    \text{h.c.})
    - (s\leftrightarrow p_i)
    .
\end{align}
Here, $c_{\bm{r},\sigma,l}$ ($c^{\dagger}_{\bm{r},\sigma,l}$) is the annihilation (creation) operator of the electron with spin $\sigma=\uparrow,\downarrow$ and orbital $l = s,p_x,p_y$ at site $\bm{r}$.
We introduce $\bm{e}_i$ with $i = x,y$ as the unit vector in the $i$-direction, with the lattice constant $a$ set to unity. We have also set $e = \hbar = 1$ here and hereafter.
We define $\Delta$ as the energy splitting between $s$ and $p$ orbitals, while $t_s$, $t_p^{a}$, $t_p^{b}$, and $t_{sp}$ represent hopping amplitudes. 
The coupling to the electric field via the vector potential $\bm{A}(t)$ is incorporated as the Peierls phase factor.

As we have seen in Sec.~\ref{sec : OM}, in multiorbital models, Wannier orbitals themselves have the electric dipole and the orbital angular momentum. 
Here we incorporate only their onsite matrix element [i.e., drop $\bm{R}_i\neq0$ terms in Eqs.~(\ref{eq : Wannier polarization}), (\ref{eq:angular-momentum-wannier})], and introduce the operator for the orbital angular momentum $L^z$ and the electric dipole $P^x,P^y$ as
\begin{align}
    L^z &= \sum_{\bm{r},\sigma}i(
    c^{\dagger}_{\bm{r},\sigma,p_y}c_{\bm{r},\sigma,p_x} - c^{\dagger}_{\bm{r},\sigma,p_x}c_{\bm{r},\sigma,p_y} ),\label{eq:Lz-operator}\\
    P^i &= P\sum_{\bm{r},\sigma}(
    c^{\dagger}_{\bm{r},\sigma,s}c_{\bm{r},\sigma,p_i} + c^{\dagger}_{\bm{r},\sigma,p_i}c_{\bm{r},\sigma,s} )
\end{align}
with
\begin{align}
    P = -e \int d^3\bm{r}w_{p_x}^{\ast}(\bm{r})xw_{s}(\bm{r}) = -e \int d^3\bm{r}w_{p_y}^{\ast}(\bm{r})yw_{s}(\bm{r}).
\end{align}
With these expressions and that for the spin angular momentum 
\begin{equation}
    S^z = \frac{1}{2}\sum_{\bm{r},\sigma,l}\sigma c^{\dagger}_{\bm{r},\sigma,l}c_{\bm{r},\sigma,l},
\end{equation}
the spin-orbit coupling $H_{\text{SOC}}$ and the electric-dipole transition term $H_{\text{dipole}}$ due to the applied electric field $\bm{E}(t)$ are written as
\begin{align}
H_{\text{SOC}} &= \frac{\lambda}{2} \sum_{\bm{r},\sigma}i\sigma \left(c^{\dagger}_{\bm{r},\sigma,p_y}c_{\bm{r},\sigma,p_x}-c^{\dagger}_{\bm{r},\sigma,p_x}c_{\bm{r},\sigma,p_y}\right),\\
    H_{\text{dipole}} &= -P\sum_{\bm{r},i,\sigma}E_i(t)(c^{\dagger}_{\bm{r},\sigma,p_i}c_{\bm{r},\sigma,s} + c^{\dagger}_{\bm{r},\sigma,s}c_{\bm{r},\sigma,p_i}).
\end{align}
Here the sign factor $\sigma$ takes $+1$ for spin-up and $-1$ for spin-down states. 

The momentum-space representation of the Hamiltonian,
\begin{align}
    H &= \sum_{\bm{k},\sigma}
        \bm{c}^{\dagger}_{\bm{k},\sigma}H_{\bm{k},\sigma}
        (t)\bm{c}_{\bm{k},\sigma}
\end{align}
with $\bm{c}^{\dagger}_{\bm{k},\sigma} = (c^{\dagger}_{\bm{k},\sigma,s}, c^{\dagger}_{\bm{k},\sigma,p_x}, c^{\dagger}_{\bm{k},\sigma,p_y})$,
can be obtained through the Fourier transformation 
$c_{\bm{r},\sigma,l} = N^{-1/2}\sum_{\bm{k}}e^{i\bm{k}\cdot\bm{r}}c_{\bm{k},\sigma,l}$. 
The matrix form of $H_{\bm{k},\sigma}(t)$ reads
\begin{align}
    H_{\bm{k},\sigma}(t) &= 
    \begin{pmatrix}
        E^s(t) & V^x(t) & V^y(t) \\
        (V^{x}(t))^{*} & \Delta + E^x(t) & -\frac{i}{2}\lambda\sigma \\
        (V^{y}(t))^{*} & \frac{i}{2}\lambda\sigma & \Delta + E^y(t)
    \end{pmatrix},
\end{align}
where the each matrix element is given by
\begin{subequations}
\begin{align}
    E^s(t) &= - 2t_s\left(\cos (k_x + A_x(t)) + \cos (k_y+A_y(t))\right), \\
    E^{x/y}(t) &= - 2t_{p}^{a/b}\cos (k_x + A_x(t)) - 2t_{p}^{b/a}\cos (k_y+A_y(t)), \\
    V^i(t) &= -2it_{sp}\sin (k_i + A_i(t)) - PE_i(t). 
\end{align}\label{eq:matrix-elements}
\end{subequations}

\begin{figure}[t]
    \centering
    \includegraphics[width = \linewidth]{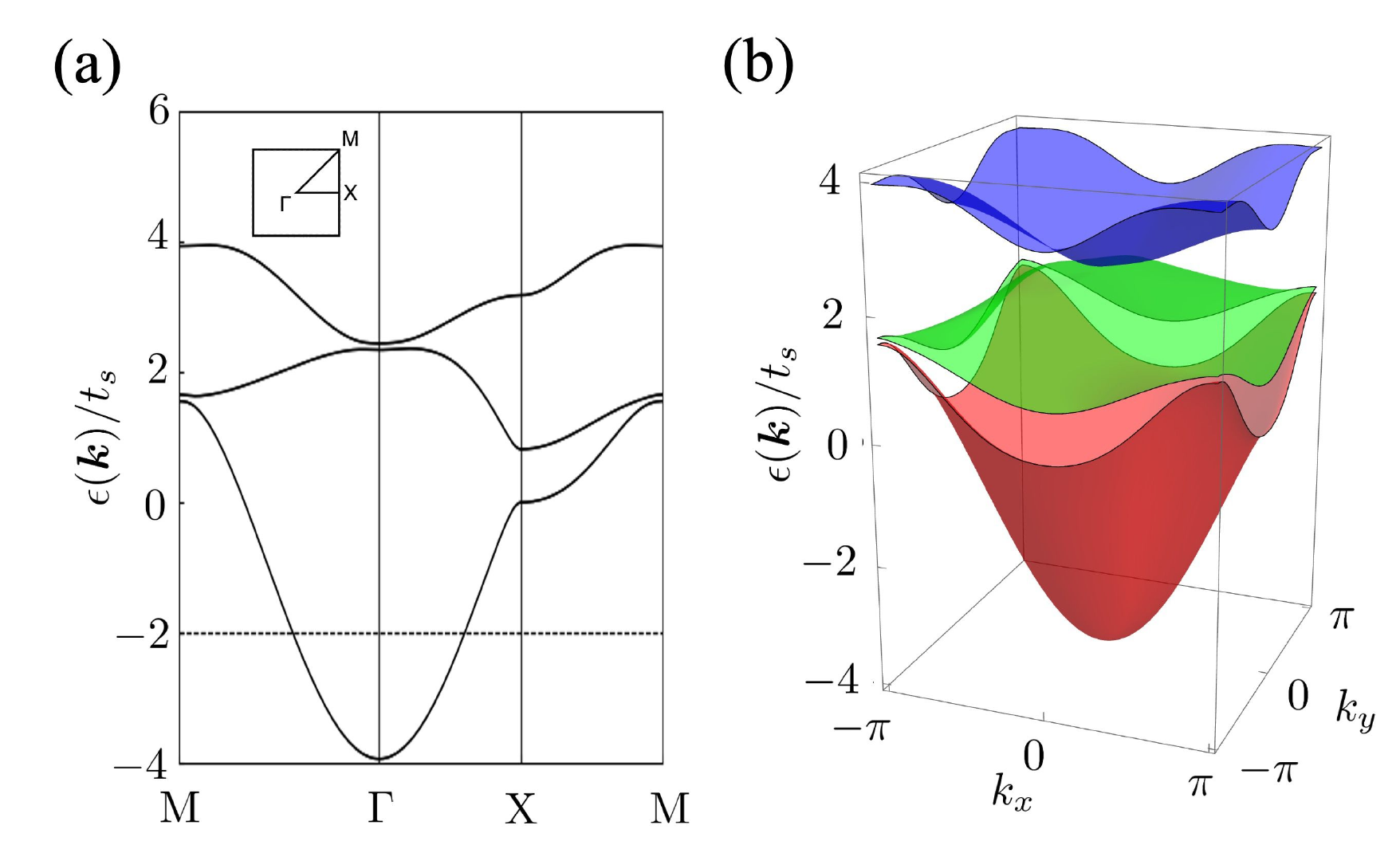}
    \caption{Band structure without an applied electric field with $\lambda = 0.1 t_s$, $\Delta = 2.0 t_s$. (a) The band structure along the high symmetry line. The dashed line represents the chemical potential $\mu = -2.0 t_s$ used in Figs.~\ref{fig: EOmega_heatmap} and \ref{fig: DeltaOmega_heatmap}. (b) 3D plot of the band structure in the first Brillouin zone.} 
    \label{fig: Band}
\end{figure}

Throughout this paper, we fix the parameters as $t_p^a = -0.4t_s, t_p^b = 0.2t_s, t_{sp} = -0.8t_s$, and $P = -1.0$. Figure~\ref{fig: Band} shows the band structure without an applied electric field. 

In this study, we focus on the circularly-polarized light represented by the vector potential $\bm{A}(t) = (A_x(t),A_y(t)) = A(-\sin \Omega t,\cos \Omega t)$. Corresponding electric field is given by $\bm{E}(t) = -\partial_t \bm{A}(t) = E(\cos \Omega t, \sin \Omega t)$ with $E = \Omega A$. 
The explicit form of the matrix elements Eq.~(\ref{eq:matrix-elements}) in the frequency domain, e.g., $E^s_m=T^{-1}\int dt E^s(t)e^{im\Omega t}$, is provided in Appendix~\ref{sec:appendix-FloquetHamiltonian}.
We note that the magnetic field is neglected in this study. As we discuss in Appendix~\ref{sec:appendix-magnetic-field}, the effect of the magnetic field on the IFE is estimated to be small compared with that of the electric field.



\subsection{Floquet formalism for nonequilibrium steady states}\label{subsec : Floquet formalism}

In this paper, we investigate the IFE in the $s$-$p$ tight-binding model introduced above by employing the Floquet formalism~\cite{Oka2019,Morimoto2023,Eckardt2017,Torre2021} and calculating the magnetic moment in the nonequilibrium steady state~\cite{Aoki2014,Seetharam2015,Morimoto2016}.

Since the present problem is time-periodic, i.e., the Hamiltonian satisfies $H(t+T) = H(t)$, we can write the solution of the time-dependent Schr\"odinger equation
\begin{align}
    i\frac{d}{dt}\ket{\psi(t)} = H(t)\ket{\psi(t)}
\end{align}
in the form of the Floquet state, as
\begin{align}
    \ket{\psi(t)} = e^{-i\epsilon_{\alpha}t}\ket{u_{\alpha}(t)}.
\end{align}
Here, $\epsilon_{\alpha}$ is the quasienergy for Floquet state $\alpha$, while $\ket{u_{\alpha}(t)}$ is the time-periodic part of the wave function, $\ket{u_{\alpha}(t+T)} = \ket{u_{\alpha}(t)}$. 
We can expand the Hamiltonian as well as the wave function
 in terms of the Floquet modes as
\begin{align}
    H(t) &= \sum_n H_{n} e^{-in\Omega t}, \\
    \ket{u_{\alpha}(t)}
    &= \sum_n \ket{u_{n,\alpha}} e^{-in\Omega t},
\end{align}
with which the time-dependent Schr\"odinger equation can be mapped to a static eigenvalue problem in the extended Hilbert space (Sambe space),
\begin{align}
    \begin{pmatrix}
        \ddots & \ddots & \ddots & & \\
        \ddots & H_{0} - \Omega & H_{1} & H_{2} & & \\
        \ddots & H_{-1} & H_{0} & H_{1} & \ddots \\
         & H_{-2} & H_{-1} & H_{0}+\Omega & \ddots \\ 
         & & \ddots & \ddots & \ddots
    \end{pmatrix}
    \begin{pmatrix}
        \vdots \\
        \vphantom{\ddots}\ket{u_{1,\alpha}} \\
        \vphantom{\ddots}\ket{u_{0,\alpha}} \\
        \vphantom{\ddots}\ket{u_{-1,\alpha}} \\
        \vdots
    \end{pmatrix}=
    \epsilon_\alpha
    \begin{pmatrix}
        \vdots \\
        \vphantom{\ddots}\ket{u_{1,\alpha}} \\
        \vphantom{\ddots}\ket{u_{0,\alpha}} \\
        \vphantom{\ddots}\ket{u_{-1,\alpha}} \\
        \vdots
    \end{pmatrix}.\label{eq:sambe}
\end{align}
We can obtain the Floquet states by diagonalizing the Floquet Hamiltonian (\ref{eq:sambe}). While there are replica states with quasienergy $\epsilon_\alpha+m\Omega$ which represent the identical state in the time domain, we restrict the range of the quasienergy to $\epsilon_\alpha\in(-\Omega/2,\Omega/2]$ to eliminate such the states.

To obtain the statistical average, we need to identify the distribution function for the Floquet states. 
Here we consider a fermionic reservoir coupled to each site, and consider the nonequilibrium steady state in the presence of dissipation to the reservoir. As we briefly reviewed in Appendix~\ref{sec:appendix-distribution}, with the nonequilibrium Green's function approach, the distribution function for the Floquet state is obtained as
\begin{align}
    f_{\alpha} = \sum_{m=-\infty}^{\infty}f(\epsilon_{\alpha}+m\Omega)\braket{u_{m,\alpha}|u_{m,\alpha}} \label{eq: Floquet distribution}
\end{align}
with $f(\epsilon) = (e^{(\epsilon-\mu)/k_BT}+1)^{-1}$ being the Fermi distribution function with the temperature $T$ and the chemical potential $\mu$ of the reservoir. 

Using Eq.~(\ref{eq: Floquet distribution}), we can evaluate the time-averaged expectation value of the operator $\mathcal{O}(t)$ as
\begin{align}
    \overline{\braket{\mathcal{O}(t)}} &= \sum_{\bm{k},\sigma,\alpha}f_{\alpha}
    \overline{\braket{u_{\alpha}(t)|\mathcal{O}(t)|u_{\alpha}(t)}}
     \nonumber\\&= \sum_{\bm{k},\sigma,\alpha}f_{\alpha}
    \sum_{m,n=-\infty}^{\infty}\braket{u_{m,\alpha}|\mathcal{O}_{mn}
    |u_{n,\alpha}}, \label{eq: Floquet expectation value}
\end{align}
where we have introduced the Sambe-space representation of operators  $\mathcal{O}_{mn} = T^{-1}\int_0^T dt \mathcal{O}(t)e^{i(m-n)\Omega t}$ and the time average $\overline{\mathcal{O}(t)} = T^{-1}\int_0^T dt \mathcal{O}(t)$.

While we are interested in calculating the orbital magnetic moment,
as shown in Eq.~(\ref{eq : lattice-contribution}), we cannot directly use the above equation to the kinetic contribution $M^z_{\text{lat}}$, since it is not expressed as an expectation value of an operator. 
We employ the following expression
\begin{multline}
    \overline{\braket{M^z_{\text{lat}}(t)}} = \frac{ie}{2\hbar}\sum_{\bm{k},\alpha}f_{\alpha}
    \overline{\braket{\partial_{k_x}u_{\alpha}(t)|(\epsilon_{\alpha}-H(t)+i\partial_t)|\partial_{k_y}u_{\alpha}(t)}}\\-(x\leftrightarrow y).
    \label{eq : Floquet expectation value for lattice contribution}
\end{multline}
by extending the first term of Eq.~(\ref{eq : lattice-contribution}) to Floquet states. The above expression is obtained from a Fourier transform of the conventional expression for the orbital magnetic moment in Sambe space.
Note that the above expression does not include the Berry curvature term [the second term of Eq.~(\ref{eq : lattice-contribution})] that arises from the change in the grand potential due to variations of the phase-space volume in the semiclassical dynamics of Bloch electrons in equilibrium systems \cite{Xiao2005}. Since the grand potential is not always well-defined in nonequilibrium systems, we do not consider this term in the present analysis.

\section{Analytical Results}\label{sec : Analytical Result}

To obtain an analytical insight, here we perform a perturbative expansion using the time-dependent Schrieffer-Wolff transformation~\cite{Bukov2016,Kitamura2017,Schrieffer1966}.
Here we consider the situation where the $s$-$p$ splitting $\Delta$ is large, and the off-diagonal elements between the $s$ and $p$ orbitals can be treated as perturbations. 
We focus on the metallic state where the Fermi level lies on the $s$-orbital band, which is expected to exhibit no magneto-optical effects when the contribution from the $p$ orbital is absent.
As we see below, hybridization with $p$ orbital leads to the emergence of the magnetic moment both for spin and orbital angular momenta, which is revealed by the calculation of the effective Hamiltonian, the effective angular-momentum operator, and the effective orbital magnetic moment operator from Wannier polarization contribution for the $s$ orbital.
As the details of the calculation are shown in Appendix~\ref{sec:appendix-SW}, here we focus on the sketch of the formalism and the highlights of the results obtained.

\subsection{Effective Hamiltonian} 
The time-dependent Schrieffer-Wolff transformation $S(t)$ is defined by the unitary transformation that generates the transformed Hamiltonian
\begin{align}
    H_{\text{SW}}(t) &= e^{iS(t)}\left(H(t)-i\frac{\partial}{\partial t}\right)e^{-iS(t)}
\end{align}
which is block-diagonal between the $s$ and $p$ orbital subspaces. 
The transformation $S(t)$ is taken time-periodic, and determined such that the offdiagonal component vanishes order by order.

As we are interested in static magnetization, let us concentrate on the time average (zeroth Fourier component) here. The time average of the $1\times1$ effective Hamiltonian for the $s$ orbital band, $\overline{H_{\text{eff}}(t)}=\mathcal{P}\overline{H_{\text{SW}}(t)}\mathcal{P}$ with $\mathcal{P}=\text{diag}(1,0,0)$, is obtained as
\begin{align}
    \overline{H_{\text{eff}}(t)}
    &= E^s_0 - \sum_m V_m(E^p_0 - (E^s_0-m\Omega)I_{2\times2})^{-1}V_m^{\dagger}
    \label{eq :  Heff}
\end{align}
at the second order of the perturbative expansion, where
\begin{align}
    V_m
    = 
    \begin{pmatrix}
        V^x_m & V^y_m
    \end{pmatrix},\quad
    E^p_0
    = 
    \begin{pmatrix}
        \Delta + E^x_0 & -\frac{i}{2}\lambda\sigma \\
        \frac{i}{2}\lambda\sigma & \Delta + E^y_0
    \end{pmatrix}
\end{align}
are the $m$th and zeroth Fourier component of the block matrices, respectively.

As the present effective Hamiltonian is $1\times1$, the effective ($k$-dependent) Zeeman field, $B_{\text{eff}}$, can be obtained as $B_{\text{eff}}=\overline{H_{\text{eff}}(t)}|_{\sigma=\uparrow}-\overline{H_{\text{eff}}(t)}|_{\sigma=\downarrow}$. 
Its leading-order expression is obtained as 
\begin{multline}
    B_{\text{eff}}
    \simeq
     \frac{E^2\lambda}{2\Omega^2}\Bigg[
        \frac{(2t_{sp}\cos k_x - P\Omega)(2t_{sp}\cos k_y - P\Omega)}{(\Delta_{x}(\bm{k})-\Omega)(\Delta_{y}(\bm{k})-\Omega)}
     \\
    - 
        \frac{(2t_{sp}\cos k_x + P\Omega)(2t_{sp}\cos k_y + P\Omega)}{(\Delta_{x}(\bm{k})+\Omega)(\Delta_{y}(\bm{k})+\Omega)}\Bigg],
    \label{eq: analytical Heff}
\end{multline}
where we define $\Delta_{x}(\bm{k})$ and $\Delta_{y}(\bm{k})$ by 
\begin{align}
    \Delta_{x/y}(\bm{k}) &= \Delta - 2(t_p^a \cos k_{x/y} + t_p^b \cos k_{y/x}) + 2t_s(\cos k_x + \cos k_y) 
\end{align}
and neglect higher-order terms $O(E^3), O(\lambda^2)$. 


It is noteworthy that the effective magnetic field $B_{\text{eff}}$ is proportional to the first power of the spin-orbit coupling constant $\lambda$. This contrasts with previous studies on the electron system with Rashba spin-orbit coupling \cite{Tanaka2020,Tanaka2024}, where the effective magnetic field is proportional to the square of the Rashba parameter. Consequently, our mechanism implies that a significant response can be observed even in materials with smaller spin-orbit coupling.

\subsection{Effective orbital angular momentum operator}

The Schrieffer-Wolff transformation also yields the perturbative expression for the orbital angular momentum as well as the spin splitting, with the construction of effective operators.
In the transformed representation under $S(t)$, the orbital angular momentum operator within the $s$ orbital subspace is written as
\begin{align}
    L^{z}_{\text{eff}}(t) &= \mathcal{P}e^{iS(t)}L^z e^{-iS(t)}\mathcal{P},
\end{align}
where $L^z$ is the matrix representation of Eq.~(\ref{eq:Lz-operator}) in the momentum space,
\begin{align}
    L^z = 
    \begin{pmatrix}
        0 & 0 & 0 \\
        0 & 0 & -i \\
        0 & i & 0
    \end{pmatrix}.
    \label{eq : Lz_matrix}
\end{align}
The time average of $L^z_{\text{eff}}(t)$ up to the second-order perturbation is given by
\begin{align}
    \overline{L^z_{\text{eff}}(t)} 
    = \sum_m& V_m(E^p_0 - (E^s_0-m\Omega)I_{2\times2})^{-1}L^z\n&\times(E^p_0 - (E^s_0-m\Omega)I_{2\times2})^{-1}V_m^{\dagger}. \label{eq: Lzeff}
\end{align}
In particular, the leading order expression reads
\begin{multline}
    \overline{L^z_{\text{eff}}(t)}
    \simeq 
    \frac{E^2}{2\Omega^2}\Bigg[\frac{(2t_{sp}\cos k_x - P\Omega)(2t_{sp}\cos k_y - P\Omega)}{(\Delta_{x}(\bm{k}) - \Omega)(\Delta_{y}(\bm{k}) - \Omega)} \\
    - \frac{(2t_{sp}\cos k_x + P\Omega)(2t_{sp}\cos k_y + P\Omega)}{(\Delta_{x}(\bm{k}) + \Omega)(\Delta_{y}(\bm{k}) + \Omega)}\Bigg],
    \label{eq: analytical Lz}
\end{multline}
where we neglect higher-order terms $O(E^3), O(\lambda)$.
As we can see in Eq.~(\ref{eq : Lz_matrix}), the time average of the orbital angular momentum operator $L^z$ for the $s$ orbital is exactly zero, but $\overline{L^z_{\text{eff}}(t)}$ is finite, which implies that the orbital angular momentum is also induced in the $s$ orbital due to the second-order perturbation between the $s$ and $p$ orbitals. 
We note that the above leading-order expression is independent of $\lambda$ and $\sigma$, unlike the time-averaged effective Hamiltonian for the $s$ orbital. This indicates that the orbital angular momentum is induced even without spin-orbit coupling.

\subsection{Effective orbital magnetic moment operator from Wannier polarization contribution}

As in the case of the effective orbital angular momentum discussed in the previous subsection, we can also obtain the effective orbital magnetic moment from the Wannier polarization contribution in the same manner.

The operator for the orbital magnetic moment from the Wannier polarization contribution is
\begin{align}
    M^z_{P}(t) 
    &= \frac{1}{4}
    \left(
        P_x \frac{\partial H(t)}{\partial k_y} - P_y \frac{\partial H(t)}{\partial k_x}
    \right)+\text{h.c.} \n
    &=\frac{P}{4}
    \begin{pmatrix}
        0 & \partial_{k_y}E^x(t) & -\partial_{k_x}E^y(t) \\
        \partial_{k_y}E^s(t) & 0 & \partial_{k_y}V^y(t) \\
        -\partial_{k_x}E^s(t) & -\partial_{k_x}V^x(t) & 0
    \end{pmatrix}
    +\text{h.c.}
\end{align}
and we consider the transformed and projected operator written as
\begin{align}
    M^z_{P,\text{eff}}(t) = \mathcal{P}e^{iS(t)}M^z_{P}(t)e^{-iS(t)}\mathcal{P}.
\end{align}
The time average of $M^z_{P,\text{eff}}(t)$ up to the first-order perturbation is obtained as
\begin{align}
    \overline{M^z_{P,\text{eff}}(t)}
    = - \sum_{m}&\Big[V_m (E^{p}_{0}-(E^{s}_{0}-m\Omega)I_{2\times 2})^{-1} W_m^{\dagger} \n
    &+ W_{-m}(E^{p}_{0}-(E^{s}_{0}+m\Omega)I_{2\times 2})^{-1}V_{-m}^{\dagger}\Big],
    \label{eq: MzPeff}
\end{align}
where $W_m$ represent the $m$th Fourier components of 
$W(t) = \mathcal{P}M^z_{P}(t)(1-\mathcal{P})$.
The leading order expression is
\begin{multline}
    \overline{M^z_{P,\text{eff}}(t)}
    \simeq 
    \frac{PE^2(t^p_b+t_s)}{4\Omega^2}
    \Bigg[\frac{(2t_{sp}\cos k_x + P\Omega)\cos k_y}{\Delta_{x}(\bm{k}) + \Omega} \\
    - \frac{(2t_{sp}\cos k_y - P\Omega)\cos k_x}{\Delta_{x}(\bm{k}) - \Omega}\Bigg] + (x \leftrightarrow y),
    \label{eq: analytical MzP}
\end{multline}
where we neglect higher-order terms $O(E^3), O(\lambda)$.

The orbital magnetic moment operator from the Wannier polarization contribution for the $s$ orbital is exactly zero.
This implies that, similar to the spin and orbital angular momentum, the transition between the $s$ orbital and the $p$ orbitals is essential for $\overline{M^z_{P,\text{eff}}(t)}$ to be nonzero.
Since the expression does not depend on $\lambda$ or $\sigma$, the Wannier polarization contribution is induced even in the absence of spin-orbit coupling, just as in the orbital angular momentum.

Let us briefly comment on the differences between the spin/orbital angular momentum contributions and the Wannier polarization contribution. First, the electric dipole $P$ is explicitly required for the Wannier polarization contribution, whereas it is not necessary for the spin or orbital angular momentum. 
Additionally, the dependence on $\Delta-m\Omega$ differs in order compared with the orbital angular momentum. 
This distinction arises because, unlike $L^z$, $M^z_{P}(t)$ inherently contains offdiagonal components, leading to a different structure in its perturbative expansion.

Finally, we note the frequency dependence of the induced magnetic moment in the low-frequency region. In the analytical expressions, Eqs.~(\ref{eq: analytical Heff}), (\ref{eq: analytical Lz}), and (\ref{eq: analytical MzP}), the dependence on $\Omega$ is proportional to $\Omega^{-1}$ as far as the perturbative expansion is valid. 
It should be noted that the perturbative expansion is valid when $E/\Omega$ is sufficiently small, due to the form of the Bessel function $\mathcal{J}(A)$ with $A = E/\Omega$ in the Floquet Hamiltonian (see Appendix~\ref{sec:appendix-FloquetHamiltonian}). This indicates that the IFE increases at low frequencies with $\propto 1/\Omega$ when $E$ is sufficiently small compared to $\Omega$.


\begin{figure*}[t]
    \centering
    \includegraphics[width = \linewidth]{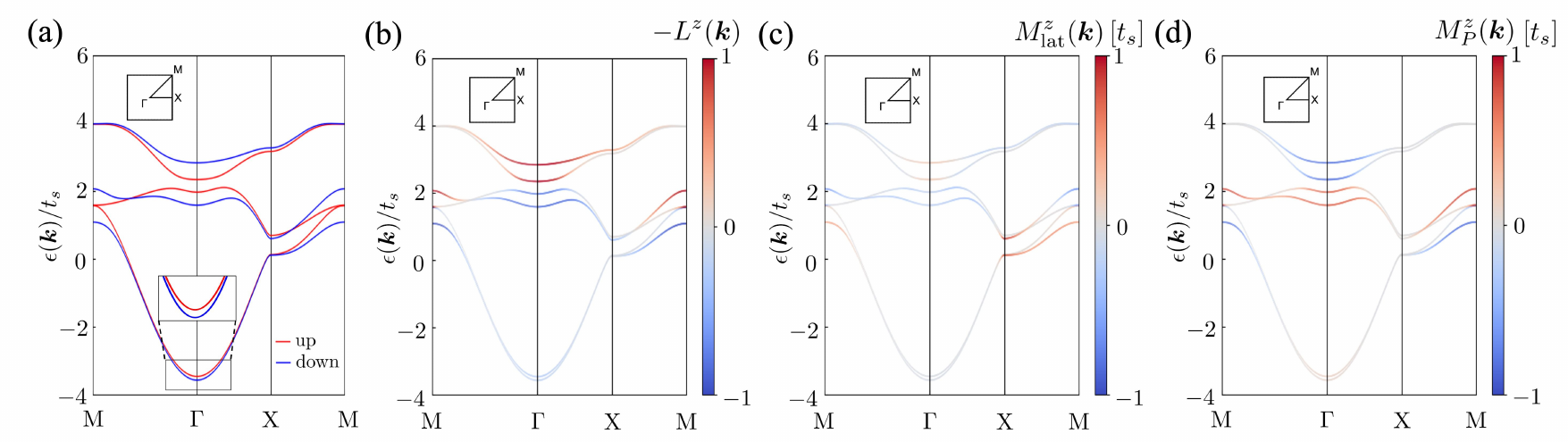}
    \caption{Floquet band structure under an applied electric field along the high symmetry line. The parameters are set to $\lambda = 0.5 t_s, \Delta = 2.0 t_s, \Omega = 8.0 t_s, E = 2.0 t_s / a$. (a) The red and blue bands correspond to the spin-up band and spin-down band, respectively. (b-d) The color represents the magnitude of the (b) orbital angular momentum $-L^{z}(\bm{k})$, (c) kinetic contribution $M^z_{\text{lat}}(\bm{k})$, and (d) Wannier polarization contribution $M^z_{P}(\bm{k})$.}
    \label{fig: Floquet band}
\end{figure*}

\begin{figure*}[t]
    \centering
    \includegraphics[width = \linewidth]{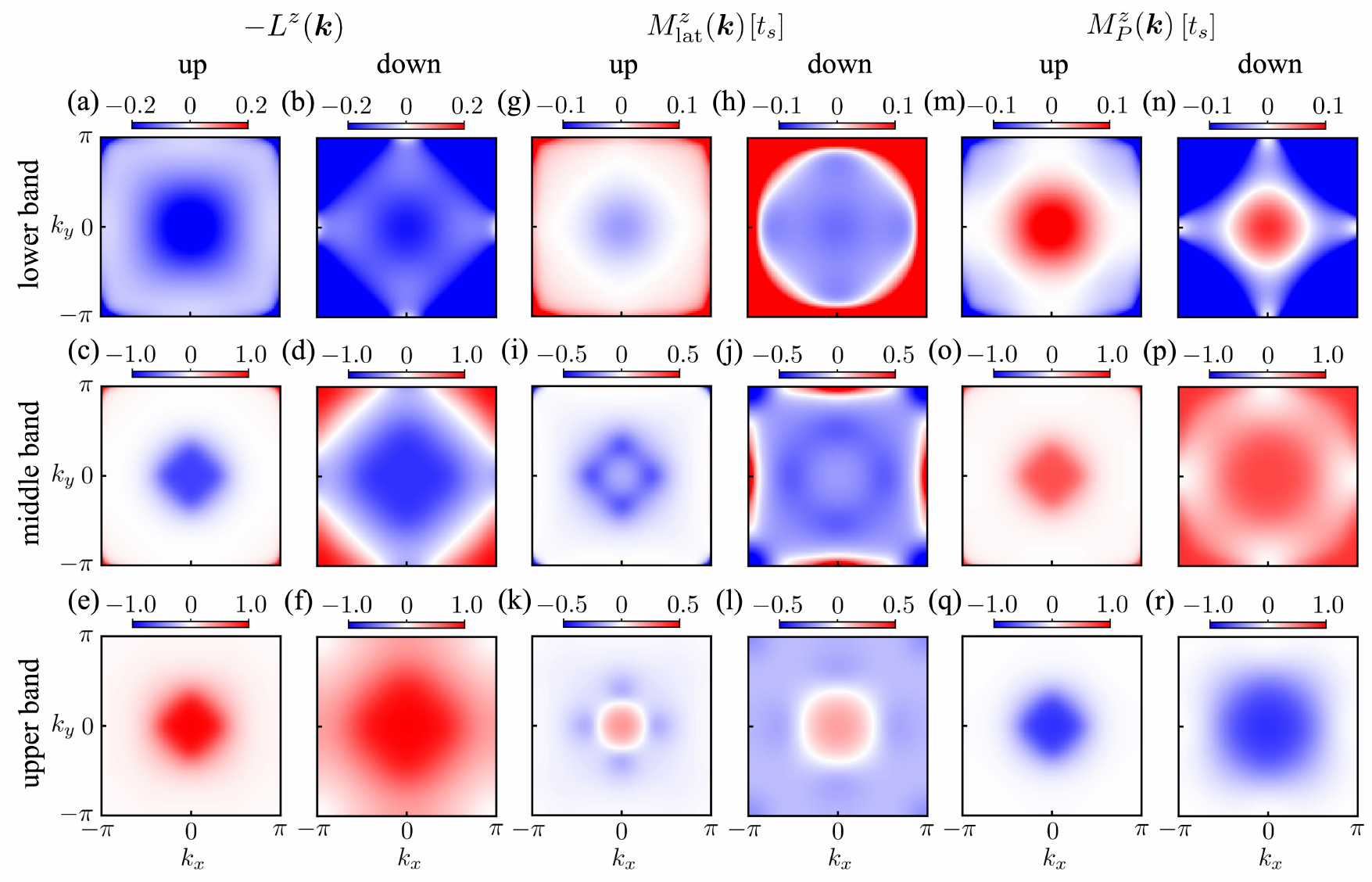}
    \caption{Distribution of (a-f) orbital angular momentum $-L^z(\bm{k})$, (g-l) kinetic contribution $M^z_{\text{lat}}(\bm{k})$, and (m-r) Wannier polarization contribution $M^z_{P}(\bm{k})$ in the first Brillouin zone for each band. The parameters are the same as Fig.~\ref{fig: Floquet band}. (a,g,m) lower band, up spin; (b,h,n) lower band, down spin; (c,i,o) middle band, up spin; (d,j,p) middle band, down spin; (e,k,q) higher band, up spin; (f,l,r) higher band, down spin. We note that, for the lower band, the color bar range is narrowed to clearly show the distribution near the $\Gamma$ point.}
    \label{fig: distribution}
\end{figure*}

\begin{figure}[t]
    \centering
    \includegraphics[width = \linewidth]{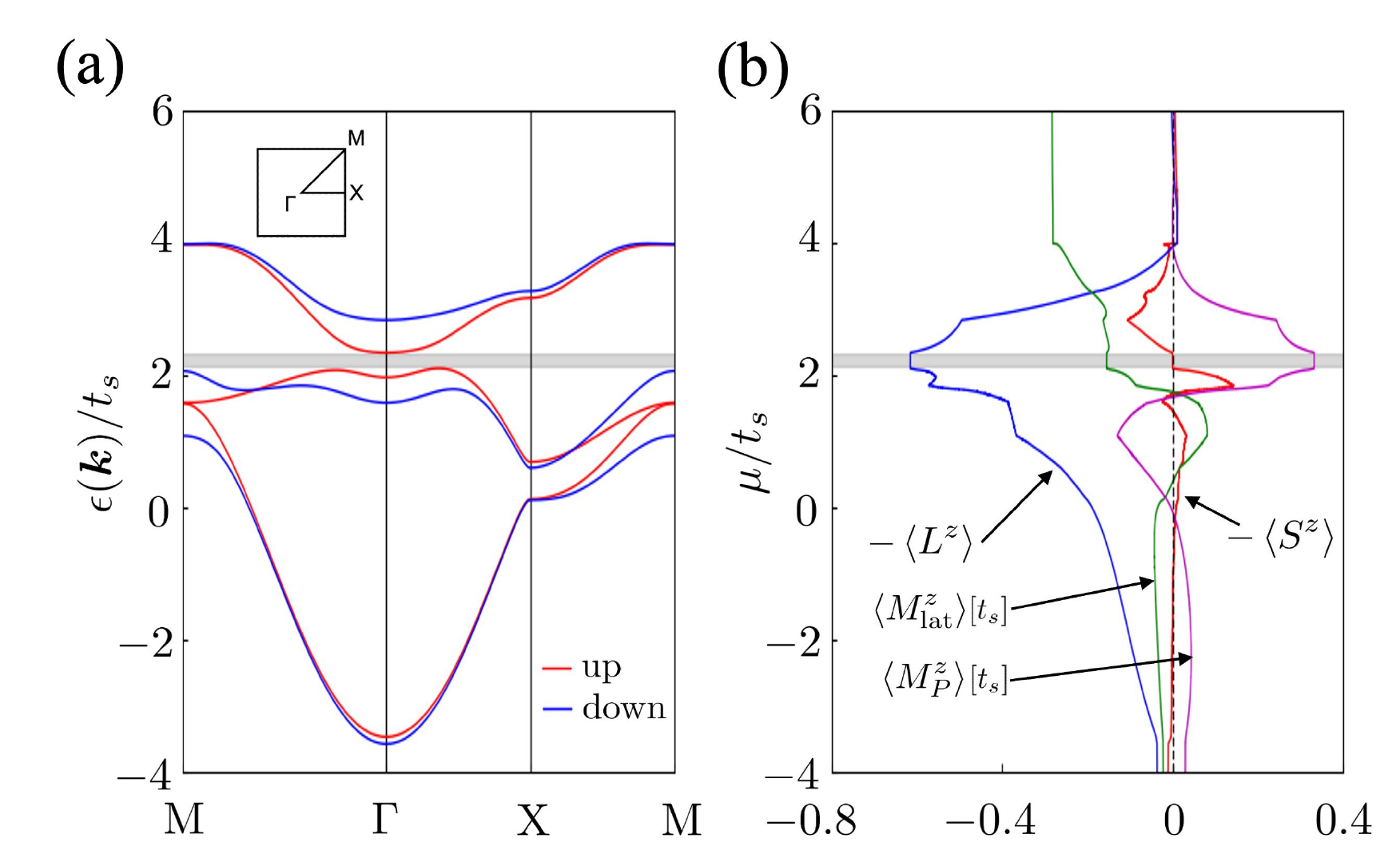}
    \caption{(a) Floquet band structure under an applied electric field along the high symmetry line. For comparison with panel (b), the figure from Fig.~\ref{fig: Floquet band}(a) is reproduced here. The shaded regions indicate the energy gap. (b) Dependence of each magnetization contribution on the chemical potential of the reservoir at zero temperature. The parameters are the same as Fig.~\ref{fig: Floquet band}.}
    \label{fig: mu_dependence}
\end{figure}

\section{Numerical Results}\label{sec : Numerical Result}

In this section, we show the numerical results of the IFE, which is obtained by diagonalizing the truncated Floquet Hamiltonian and calculating the statistical average using Eqs.~(\ref{eq: Floquet distribution})-(\ref{eq : Floquet expectation value for lattice contribution}).

\subsection{Floquet band structure}

Figure~\ref{fig: Floquet band}(a) shows the Floquet band structure with $\Delta = 2.0 t_s, \Omega = 8.0 t_s, E = 2.0 t_s$, where the photon energy is large enough that the mixing between the original bands and the replica bands is small. 
Here we consider a relatively large spin-orbit coupling of $\lambda = 0.5 t_s$.
Here we only show the bands $\epsilon_\alpha\in(-\Omega/2,\Omega/2]$, while the replica bands are omitted.

Let us focus on the area around the $\Gamma$ point of the lower band originating from the $s$ orbital band.
In Fig.~\ref{fig: Floquet band}(a) we can see the splitting into spin-up and spin-down components around the $\Gamma$ point in the lower band, which indicates that, as we see in Eq.~(\ref{eq: analytical Heff}) in the analytical calculations, spin splitting indeed occurs even in the $s$ orbital.

Figures~\ref{fig: Floquet band}(b,c,d) illustrate the same Floquet bands, with the color indicating the values of the orbital angular momentum $-L^z(\bm{k})$, kinetic contribution $M^z_{\text{lat}}(\bm{k})$, and Wannier polarization contribution $M^z_{P}(\bm{k})$, respectively. Figure~\ref{fig: distribution} provides a detailed view of the distribution in the first Brillouin zone for each band, corresponding to Fig.~\ref{fig: Floquet band}(b,c,d). Near the $\Gamma$ point in the lower band, it is evident that all these contributions share the same sign for the spin-up and spin-down bands. This observation aligns with the analytical result in Eq.~(\ref{eq: analytical Lz}), which shows that $L^z(\bm{k})$ is independent of the spin degree of freedom.

\subsection{Fermi surface contribution and Fermi sea contribution}

Figure~\ref{fig: mu_dependence}(b) illustrates how the expectation values of each contribution depend on the chemical potential of the reservoir, with the parameters the same as Fig.~\ref{fig: Floquet band}. The shaded regions represent areas where the energy gap is open. As the chemical potential increases from $\mu = -4t_s$, the spin magnetic moment remains nearly constant. On the other hand, the orbital magnetic moment from the three contributions tends to increase in magnitude, although the kinetic contribution and the Wannier polarization contribution undergo a sign change.
When the chemical potential reaches the energy gap region, the spin magnetic moment becomes zero, while the other contributions to the orbital magnetic moment remain nonzero.
If the electrons filled the Floquet bands from the bottom as in equilibrium, the contributions from the spin-up and spin-down bands cancel each other for the spin magnetic moment. This makes the spin magnetic moment primarily determined by contributions near the Fermi level (i.e., it is a Fermi-surface contribution). In contrast, for the orbital magnetic moment, the contributions from the spin-up and spin-down bands have the same sign, as mentioned in the previous subsection. Thus the contributions do not cancel each other, resulting in a Fermi-sea contribution. While the distribution function here differs from the equilibrium one and the classification into the Fermi-surface/sea contributions is not well-defined, qualitative behavior agrees with this interpretation, as the nonequilibirum distribution function behaves similarly to the equilibrium one for off-resonant states. The distinction discussed here suggests that the orbital magnetic moment is expected to be larger than the spin magnetic moment.

\subsection{Frequency dependence of IFE}

\begin{figure*}[t]
    \centering
    \includegraphics[width = \linewidth]{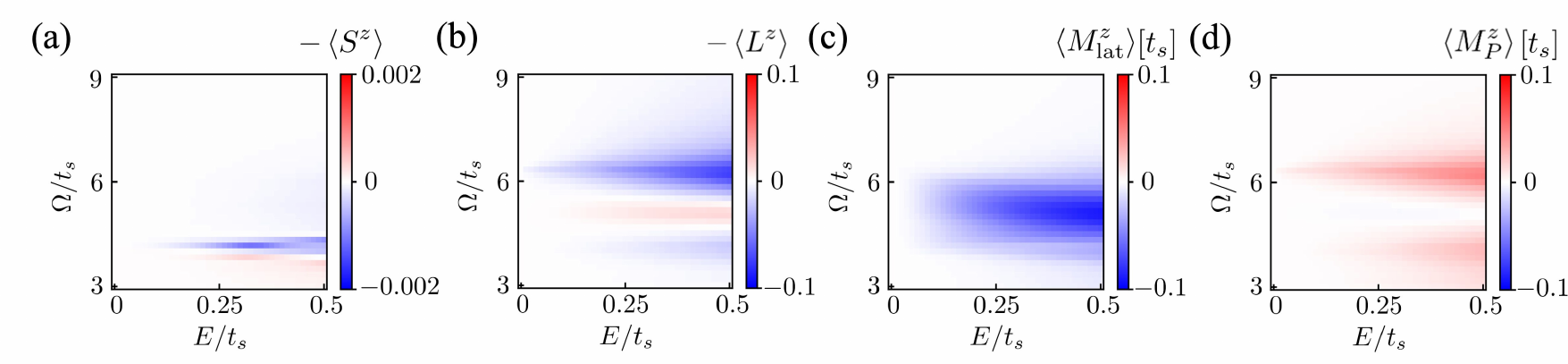}
    \caption{$E$ and $\Omega$ dependence of the (a) spin angular momentum $-\braket{S^z}$, (b) orbital angular momentum $-\braket{L^z}$, (c) kinetic contribution $\braket{M^z_{\text{lat}}}$, and (d) Wannier polarization contribution $\braket{M^z_{P}}$. The parameters are set to $\lambda = 0.1 t_s, \Delta = 2.0 t_s, k_B T = 0.01 t_s, \mu = -2.0t_s$.}
    \label{fig: EOmega_heatmap}
\end{figure*}

\begin{figure*}[t]
    \centering
    \includegraphics[width = \linewidth]{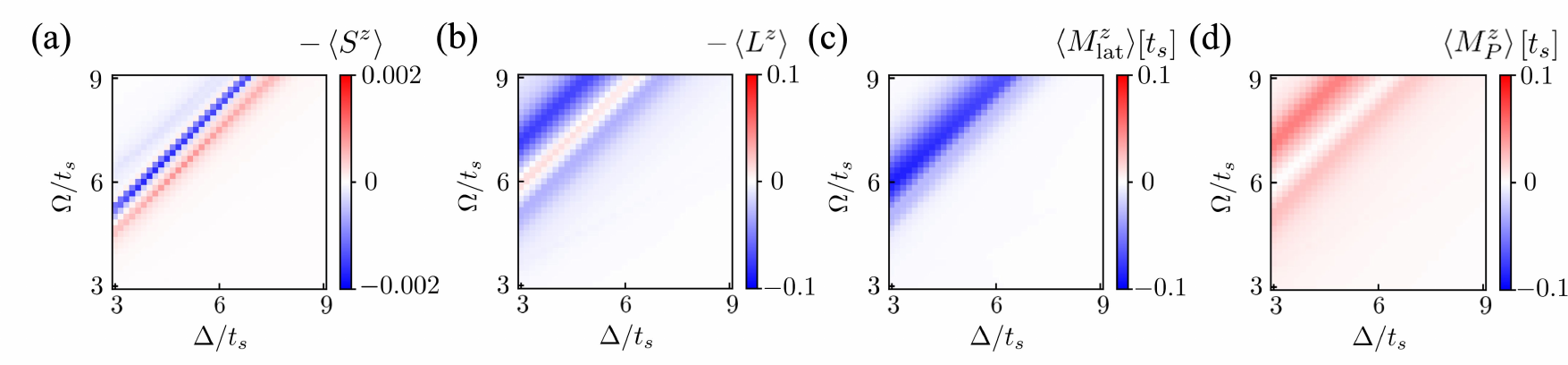}
    \caption{$\Delta$ and $\Omega$ dependence of the (a) spin angular momentum $-\braket{S^z}$, (b) orbital angular momentum $-\braket{L^z}$, (c) kinetic contribution $\braket{M^z_{\text{lat}}}$, and (d) Wannier polarization contribution $\braket{M^z_{P}}$. The parameters are set to $\lambda = 0.1 t_s, E = 0.5 t_s, k_B T = 0.01 t_s, \mu = -2.0t_s$.}
    \label{fig: DeltaOmega_heatmap}
\end{figure*}

Figures~\ref{fig: EOmega_heatmap} and~\ref{fig: DeltaOmega_heatmap} show the frequency dependence of the expected values of the four types of magnetic moment. Specifically, Fig.~\ref{fig: EOmega_heatmap} shows the electric field strength $E$ and the frequency $\Omega$ dependence while Fig.~\ref{fig: DeltaOmega_heatmap} depicts the $s$-$p$ splitting $\Delta$ and the frequency $\Omega$ dependence.
Here we turn to the case with the weak spin-orbit coupling $\lambda = 0.1 t_s$. We fix the temperature and the chemical potential of the reservoir as $k_B T = 0.01 t_s, \mu = -2.0t_s$, respectively, for both figures (See also Fig.~\ref{fig: Band}).
We set $\Delta = 2.0 t_s$ in Fig.~\ref{fig: EOmega_heatmap}, while $E = 0.5 t_s$ is adopted in Fig.~\ref{fig: DeltaOmega_heatmap}.
We varied the frequency from below the bandgap to around the bandwidth.

In Fig.~\ref{fig: EOmega_heatmap}, we can see that the response is enhanced at specific frequencies for all four types of magnetic moment: $\Omega \simeq 4t_s$ for spin angular momentum, $\Omega \simeq 4t_s, 6t_s$ for orbital angular momentum, $\Omega \simeq 5t_s$ for kinetic contribution, and $\Omega \simeq 4t_s, 6t_s$ for Wannier polarization contribution. As we can see from Fig.~\ref{fig: Floquet band}, the band gap between the $s$ orbital band and the other two bands ranges from $4t_s$ to $7t_s$ around the $\Gamma$ point.
This analysis indicates significant changes in system behavior as the frequency approaches resonance. In particular, we can see that the response of the IFE is generally enhanced near the resonance frequencies.

For the spin/orbital angular momentum and Wannier polarization contribution, we observed two peak structures as the frequency varies in Figs.~\ref{fig: EOmega_heatmap} and~\ref{fig: DeltaOmega_heatmap}. This can be understood from the analytical expressions in Eqs.~(\ref{eq: analytical Heff}), (\ref{eq: analytical Lz}), and (\ref{eq: analytical MzP}). 
These equations contain terms proportional to $[(\Delta_{x}(\bm{k}) - \Omega)(\Delta_{y}(\bm{k}) - \Omega)]^{-1}$ or $[(\Delta_{x/y}(\bm{k}) - \Omega)]^{-1}$, and the response is enhanced near the resonance frequencies, i.e., when the frequency $\Omega$ is close to $\Delta_{x}(\bm{k})$ and $\Delta_{y}(\bm{k})$. 
On the other hand, the kinetic contribution does not exhibit a double peak structure but rather shows a broader single peak structure.

From Fig.~\ref{fig: DeltaOmega_heatmap}, we can see that the resonant frequency increases linearly with the $s$-$p$ splitting $\Delta$, consistent with the interpretation mentioned above. In particular, 
the peak of the spin magnetic moment is sharper compared to that of the orbital magnetic moment, which can be interpreted as follows.
The spin magnetic moment exhibits a Fermi surface-like behavior, where only electrons within a narrow energy range near the Fermi level contribute to the response. In contrast, the orbital magnetic moment exhibits a Fermi sea-like behavior, where electrons filling the bands from the bottom, spanning a broader energy range, contribute to the response.

\section{Discussion}\label{sec : Discussion}

We have investigated the $s$-$p$ tight-binding model irradiated by circularly-polarized light as a minimal model for the IFE and investigated four types of magnetic moments in the same system. We performed perturbative analysis using time-dependent Schrieffer-Wolff transformation and numerical evaluation of magnetization based on the Floquet formalism.

Let us estimate the orders of magnitude of the magnetic moment using the peak value in Fig.~\ref{fig: DeltaOmega_heatmap}.
Assuming a lattice constant of $a = 3~\mathrm{\AA}$ and a hopping amplitude of $t_s=1$~eV, $E = 0.5t_s/a$ corresponds to an electric field of $E = 17 $~MV/cm. 
First, the contribution from the spin/orbital angular momentum in the unit of the Bohr magneton $\mu_B$ are read as $-2\mu_B\braket{S^z}/\hbar = -4 \times 10^{-3}\mu_B$, $-\mu_B\braket{L^z}/\hbar = -1\times 10^{-1}\mu_B$, indicating that the orbital angular momentum is about 25 times larger than the spin magnetization.
On the other hand, the orbital magnetic moment from kinetic and Wannier polarization contribution depend not only on the fundamental constant $\mu_B$ but also on the system's energy scale $t_s$ and the lattice constant $a$. Using $a = 3 ~\mathrm{\AA}$ and $t_s = 1$~eV, we obtain $ea^2t_s/\hbar = 2.4 \mu_B$. Thus, the orders of the magnitude are estimated as $\braket{M^z_{\text{lat}}} = -2 \times 10^{-1}\mu_B$ and $\braket{M^z_{P}} = 1 \times 10^{-1}\mu_B$, which are both larger than the spin angular momentum and comparable to the orbital angular momentum. The fact that the orbital magnetic moment is significantly larger than the spin magnetic moment can be explained by the difference in the dependence on spin-orbit interaction and the distinction between the Fermi surface contribution and the Fermi sea contribution.
In Ref.~\cite{Berritta2016}, the IFE in several materials was investigated using first-principles calculation, and the induced magnetic moment was estimated to be of the order of $10^{-3}\mu_B$-$10^{-2}\mu_B$  for an electric field of $E = 7.5$ MV/cm, which is typically one order smaller than our estimation.
Using a typical magnetic susceptibility of $\chi = 1 \times 10^{-5}$, we can estimate the effective magnetic field corresponding to the induced magnetic moment in our model. Using $\braket{M^z_{\text{tot}}} = -2 \times 10^{-1}\mu_B$, the calculation indicates an effective field of as large as $B_{\text{eff}} = 8600 $ T.
Assuming that the magnetization scales with the square of the electric field, an electric field of $E = 1$ MV/cm corresponds to a strong effective magnetic field of $B_{\text{eff}} =30$ T.

In conclusion, we found that the response of the orbital magnetic moment is larger than that of the spin magnetic moment by the order of magnitude. We also found that the magnitudes of the three types of orbital magnetization, namely, orbital angular momentum, kinetic contribution, and Wannier polarization contribution, can be comparable. These results emphasize the importance of multiorbital effects and equal-footing treatment of all contributions in the IFE. 
We hope that our study provides new insight into the physical understanding of the IFE.

\acknowledgements
We thank M. Sato for fruitful discussions.
This work was supported by
JSPS KAKENHI Grant 23K25816, 23K17665, 24H02231 (T.M.), and 25K07219 (S.K.). 
K.T. was supported by the Forefront Physics and Mathematics program to drive transformation (FoPM).

\appendix

\section{Derivation of Eq.~(\ref{eq:bloch-to-lattice})}\label{sec:appendix-wannier}

In this appendix, we derive the relation between the lattice eigenstate
and the Bloch wave function given by Eq.~(\ref{eq:bloch-to-lattice}). We consider
a periodic potential $V(\bm{r})$ satisfying $V(\bm{r})=V(\bm{r}-\bm{R}_{i})$,
with which the eigenstate of the Hamiltonian in the continuous space
can be taken as
\begin{equation}
\psi_{\alpha\bm{k}}(\bm{r})=\frac{1}{\sqrt{N}}u_{\alpha\bm{k}}(\bm{r})e^{i\bm{k}\cdot\bm{r}}
\end{equation}
with $u_{\alpha\bm{k}}(\bm{r})=u_{\alpha\bm{k}}(\bm{r}-\bm{R}_{i})$,
thanks to the Bloch theorem. Here the Hamiltonian for the periodic
part $u_{\alpha\bm{k}}$ is given by 
\begin{gather}
\mathcal{H}(\bm{k})=\frac{\hbar^{2}}{2m_{e}}(-i\bm{\nabla}+\bm{k})^{2}+V(\bm{r})
\end{gather}
with the electron mass $m_{e}$,
which satisfies $\mathcal{H}(\bm{k})u_{\alpha\bm{k}}(\bm{r})=E_{\alpha\bm{k}}u_{\alpha\bm{k}}(\bm{r})$. The maximally-localized Wannier
function (MLWF) is defined as the Fourier transform of the Bloch wave
function $\psi_{\alpha\bm{k}}$, with a $\bm{k}$-dependent unitary
matrix $U(\bm{k})$ mixing energetically close bands (so-called entangled
bands)~\cite{Marzari2012}, i.e. 
\begin{align}
w_{l}(\bm{r}-\bm{R}_{i}) & =\frac{1}{\sqrt{N}}\sideset{}{^{\prime}}\sum_{\alpha}\sum_{\bm{k}}U_{l\alpha}(\bm{k})\psi_{\alpha\bm{k}}(\bm{r})e^{-i\bm{k}\cdot\bm{R}_{i}}.\label{eq:wannier-to-bloch}
\end{align}
Here the primed sum $\sum_{\alpha}^{\prime}$ indicates
that the sum over the bands $\alpha$ are taken within the entangled
bands. The arbitrariness of the unitary matrix $U(\bm{k})$ and the
overall phase factor of $\psi_{\alpha\bm{k}}$ are used to minimize
the spread of the Wannier function. To obtain exponentially-localized
MLWFs, $U\psi$ should have no discontinuity as a function of $\bm{k}$. 

Since the MLWFs form a complete basis, the field operator in the
continuous space, $\psi(\bm{r})$, can be expanded as 
\begin{equation}
\psi(\bm{r})=\sum_{jl}w_{l}(\bm{r}-\bm{R}_{j})c_{jl}.\label{eq:field-op-wannier}
\end{equation}
Then the Hamiltonian $\mathcal{H}$ in the second-quantized form can
be expanded as 
\begin{align}
\mathcal{H} & =\int d^{3}\bm{r}\psi^{\dagger}(\bm{r})\left[-\frac{\hbar^{2}\bm{\nabla}^{2}}{2m_{e}}+V(\bm{r})\right]\psi(\bm{r})
=\sum_{ijll^\prime}c_{il}^{\dagger}H_{ll^\prime,ij}c_{jl^\prime},\label{eq:wannier-rep}
\end{align}
where 
\begin{align}
H_{ll^\prime,ij} & =\frac{1}{N}\sideset{}{^{\prime}}\sum_{\alpha}\sum_{\bm{k}}E_{\alpha\bm{k}}U_{\alpha l}^{\dagger}(\bm{k})U_{l^\prime\alpha}(\bm{k})e^{i\bm{k}\cdot(\bm{R}_{i}-\bm{R}_{j})}.
\end{align}
 The tight-binding (lattice) Hamiltonian can be obtained by restricting
the sum in Eq.~(\ref{eq:wannier-rep}) on entangled bands close to the
Fermi level.

The Fourier-transformed lattice Hamiltonian is then obtained as $H_{ll^\prime}(\bm{k})=\sum_{\alpha}^{\prime}E_{\alpha\bm{k}}U_{\alpha l}^{\dagger}(\bm{k})U_{l^\prime\alpha}(\bm{k})$, which
satisfies
\begin{equation}
\sideset{}{^{\prime}}\sum_{l^\prime}H_{ll^\prime}(\bm{k})U_{\alpha l^\prime}^{\dagger}(\bm{k})=E_{\alpha\bm{k}}U_{\alpha l}^{\dagger}(\bm{k}).
\end{equation}
This shows that the unitary matrix is written by the lattice eigenstate
$\phi_{\alpha\bm{k}l}$ as
\begin{equation}
U_{\alpha l}^\dagger(\bm{k})=\phi_{\alpha\bm{k}l},
\end{equation}
with which the inverse relation of Eq.~(\ref{eq:wannier-to-bloch}) coincides with
Eq.~(\ref{eq:bloch-to-lattice}).

\section{Derivation of the light-matter Hamiltonian including dipole interaction}\label{sec:appendix-dipole}

In this appendix, we derive the dipole interaction Eq.~(\ref{eq:dipole-interaction})
as a Wannier correction to the Peierls phase. Here we focus on homogeneous
electric fields. The dipole interaction can be straightforwardly obtained
in the length gauge, $\bm{A}(t)=0,$ $\phi(\bm{r},t)=-\bm{E}(t)\cdot\bm{r}$.
By expanding the scalar potential term into the Wannier orbital using
Eq.~(\ref{eq:field-op-wannier}), we obtain
\begin{align}
H_{E} & =\int d^{3}\bm{r}\psi^{\dagger}(\bm{r})e\bm{E}(t)\cdot\bm{r}\psi(\bm{r})\\
 & =\sum_{ijll^\prime}c_{il}^{\dagger}\left[\int d^{3}\bm{r}w_{l}^{\ast}(\bm{r}-\bm{R}_{i})e\bm{E}(t)\cdot\bm{r}w_{l^\prime}(\bm{r}-\bm{R}_{j})\right]c_{jl^\prime}\\
 & =\sum_{il}e\bm{E}(t)\cdot\bm{R}_{i}c_{il}^{\dagger}c_{il}-\sum_{ijll^\prime}c_{il}^{\dagger}(\bm{E}(t)\cdot\bm{P}_{ll^\prime,ij})c_{jl^\prime},\label{eq:scalar-potential}
\end{align}
where in the last line we have shifted the integrand as $\bm{r}\to\bm{R}_{i}+\bm{r}$,
and defined $\bm{P}_{ll^\prime,ij}=N^{-1}\sum_{\bm{k}}\bm{P}_{ll^\prime}(\bm{k})e^{i\bm{k}\cdot(\bm{R}_{i}-\bm{R}_{j})}$.
The first term represents the discretized scalar potential independent
of the orbital degree of freedom, while the second term restricted
to the entangled band is nothing but the dipole interaction due to
the Wannier correction to the electric dipole $\bm{P}$. 

On the other hand, formulation with the velocity gauge, $\bm{A}(t)=-\int^{t}dt\bm{E}(t),\phi(\bm{r},t)=0$,
is more tricky. A simple way to obtain the Peierls phase factor is
to perform the (lattice) gauge transformation of Eq.~(\ref{eq:scalar-potential}).
Since the local gauge transformation on the lattice Hamiltonian, 
\begin{equation}
c_{il}\to c_{il}e^{i\Lambda_{i}(t)}=e^{-i\sum_{il}\Lambda_{i}(t)c_{il}^{\dagger}c_{il}}c_{il}e^{i\sum_{il}\Lambda_{i}(t)c_{il}^{\dagger}c_{il}},
\end{equation}
does not alter physical observables, the transformed Hamiltonian 
\begin{equation}
H^{\prime}(t)=e^{-i\sum_{il}\Lambda_{i}(t)c_{il}^{\dagger}c_{il}}\left(H+H_{E}-i\hbar\frac{\partial}{\partial t}\right)e^{i\sum_{il}\Lambda_{i}(t)c_{il}^{\dagger}c_{il}}
\end{equation}
is equivalent to the original Hamiltonian. By choosing $\Lambda_{i}(t)=e\bm{A}(t)\cdot\bm{R}_{i}/\hbar$
with $\bm{E}(t)=-\partial_{t}\bm{A}(t)$, the hopping term is modified
as
\begin{align}
c_{il}^{\dagger}c_{jl^\prime} & \to c_{il}^{\dagger}c_{jl^\prime}e^{-i(\Lambda_{i}(t)-\Lambda_{j}(t))}\\
 & =c_{il}^{\dagger}c_{jl^\prime}e^{-ie\bm{A}(t)\cdot(\bm{R}_{i}-\bm{R}_{j})/\hbar},
\end{align}
which is nothing but the Peierls phase factor. Combined with
\begin{equation}
-i\hbar e^{-i\sum_{il}\Lambda_{i}(t)c_{il}^{\dagger}c_{il}}\frac{\partial}{\partial t}e^{i\sum_{il}\Lambda_{i}(t)c_{il}^{\dagger}c_{il}}=-\sum_{il}e\bm{E}(t)\cdot\bm{R}_{i}c_{il}^{\dagger}c_{il},
\end{equation}
we obtain
\begin{align}
H^{\prime}(t) & =\sum_{ijll^\prime}c_{il}^{\dagger}\left(H_{ll^\prime,ij}-\bm{E}(t)\cdot\bm{P}_{ll^\prime,ij}\right)c_{jl^\prime}e^{-ie\bm{A}(t)\cdot(\bm{R}_{i}-\bm{R}_{j})/\hbar},\label{eq:velocity-gauge}
\end{align}
which includes both the Peierls phase factor and the dipole interaction.

Note that, if we expand the the velocity-gauge Hamiltonian in the
continuous space 
\begin{equation}
\mathcal{H}_{\text{vel}}(t)=\frac{1}{2m_{e}}(-i\hbar\bm{\nabla}+e\bm{A}(t))^{2}+V(\bm{r})
\end{equation}
into the Wannier orbital as in Eq.~(\ref{eq:scalar-potential}), we cannot reproduce
Eq.~(\ref{eq:velocity-gauge}), and even worse, the gauge invariance of physical
observables is broken. This is because the Bloch wave function (and
Wannier orbital) in the absence of the electric field does not provide
a (long-time) solution of the velocity-gauge Hamiltonian, even in
the case of an infinitesimal field. In order to capture the correct
weak-field limit, one has to employ the adiabatic theorem and construct
the adiabatic evolution of the Bloch wave function, described by
\begin{equation}
\tilde{\psi}_{\alpha\bm{k}}(\bm{r},t)=u_{\alpha,\bm{k}+e\bm{A}(t)/\hbar}(\bm{r})e^{i\bm{k}\cdot\bm{r}}.
\end{equation}
Accordingly, the Wannier orbital should also be modified, as 
\begin{align}
\tilde{w}_{l}(\bm{r}-\bm{R}_{i},t) & =\frac{1}{N}\sideset{}{^{\prime}}\sum_{\bm{k}\alpha}\phi_{\alpha,\bm{k}+e\bm{A}(t)/\hbar,l}^{\ast}u_{\alpha,\bm{k}+e\bm{A}(t)/\hbar}(\bm{r})e^{i\bm{k}\cdot(\bm{r}-\bm{R}_{i})}.
\end{align}
The subspace spanned by these modified Wannier orbitals is indeed
suitable for extracting the low-energy Hamiltonian with external electric
fields. Let us introduce $\psi(\bm{r},t)$ as a solution of the time-dependent
Schr\"odinger equation, $i\hbar\partial_{t}\psi(\bm{r},t)=\mathcal{H}_{\text{vel}}(t)\psi(\bm{r},t)$,
and expand it by the modified Wannier function as
\begin{equation}
\psi(\bm{r},t)=\sum_{il}c_{il}(t)\tilde{w}_{l}(\bm{r}-\bm{R}_{i},t).
\end{equation}
Then the equation of motion for the coefficient $c_{il}(t)$ is derived
as
\begin{align}
i\hbar\frac{\partial c_{il}}{\partial t} & =\frac{1}{N}\sideset{}{^{\prime}}\sum_{\bm{k}jl^\prime\alpha}\left[E_{\alpha\bm{k}}\phi_{\alpha\bm{k}l}\phi_{\alpha\bm{k}l^\prime}^{\ast}\right]_{\bm{k}\to\bm{k}+e\bm{A}(t)/\hbar}c_{jl^\prime}e^{i\bm{k}\cdot(\bm{R}_{i}-\bm{R}_{j})}\nonumber\\
 & -\frac{1}{N}\sum_{\bm{k}jl^\prime}\left[\bm{E}(t)\cdot\bm{P}_{ll^\prime}(\bm{k})\right]_{\bm{k}\to\bm{k}+e\bm{A}(t)/\hbar}c_{jl^\prime}e^{i\bm{k}\cdot(\bm{R}_{i}-\bm{R}_{j})},
\end{align}
where the first term represents the instantaneous eigenenergy and results
in the dynamical phase factor, while the second term comes from the
time derivative of the modified Wannier orbital. Performing $k$ summation
after a shift of the variable $\bm{k}\to\bm{k}-e\bm{A}(t)/\hbar$
yields 
\begin{align}
i\hbar\frac{\partial c_{il}}{\partial t} & =\sum_{ijl^\prime}\left(H_{ll^\prime,ij}-\bm{E}(t)\cdot\bm{P}_{ll^\prime,ij}\right)e^{-ie\bm{A}(t)\cdot(\bm{R}_{i}-\bm{R}_{j})/\hbar}c_{jl^\prime},
\end{align}
which coincides with the time evolution under $H^{\prime}(t)$.

\section{Details of the orbital magnetization in multiorbital systems}\label{sec:appendix-OM}

In this appendix, we provide the derivation of Eq.~(\ref{eq:orbitalmag-decomp}). Here
we start with the formula for the orbital magnetization $\bm{M}_{\text{orb}}$
in terms of the Bloch wave function \cite{Shi2007}, $\bm{M}_{\text{orb}}=\bm{M}^{(1)}+\bm{M}^{(2)}$ with
\begin{align}
\bm{M}^{(1)} & =\frac{ie}{2\hbar}\sum_{\bm{k}\alpha}\int_{c}d^{3}\bm{r}\frac{\partial u_{\alpha\bm{k}}^{\ast}(\bm{r})}{\partial\bm{k}}\times(E_{\alpha\bm{k}}-\mathcal{H}(\bm{k}))\frac{\partial u_{\alpha\bm{k}}(\bm{r})}{\partial\bm{k}}f_{\alpha},\\
\bm{M}^{(2)} & =-\frac{ek_{B}T}{\hbar}\sum_{\bm{k}\alpha}\frac{\partial}{\partial\bm{k}}\times\bm{a}_{\alpha\alpha}(\bm{k})\ln(1-f_{\alpha}),
\end{align}
and decompose the contribution into the terms written by the lattice
eigenstate $\phi_{\alpha\bm{k}m}$ and the terms written as the Wannier
correction. Here $\int_{c}$ denotes the volume integral within a
unit cell, and $f_{\alpha}=f(E_{\alpha\bm{k}})$.

Let us first decompose the first term $\bm{M}^{(1)}$. By inserting
the resolution of the identity 
\begin{equation}
\sum_{\alpha}u_{\alpha\bm{k}}(\bm{r})u_{\alpha\bm{k}}^{\ast}(\bm{r}^{\prime})=\sum_{i}\delta(\bm{r}-\bm{r}^{\prime}-\bm{R}_{i}),
\end{equation}
$\bm{M}^{(1)}$ can be rewritten as 
\begin{align}
\bm{M}^{(1)} & =\frac{ie}{2\hbar}\sum_{\bm{k}\alpha}\sum_{\beta\neq\alpha}\bm{a}_{\alpha\beta}(\bm{k})\times(E_{\alpha\bm{k}}-E_{\beta\bm{k}})\bm{a}_{\beta\alpha}(\bm{k})f_{\alpha}.\label{eq:m1-dipole}
\end{align}
As we have already shown in Eq.~(\ref{eq:dipole-operator}), the electric dipole$-e\bm{a}_{\alpha\beta}$
can be decomposed into the lattice counterpart and the Wannier correction.
Using this relation to Eq.~(\ref{eq:m1-dipole}), we obtain 
\begin{align}
\bm{M}^{(1)} & =\frac{ie}{2\hbar}\sum_{\bm{k}\alpha}\sum_{\beta\neq\alpha}[\bm{a}_{\text{lat}}(\bm{k})]_{\alpha\beta}\times(E_{\alpha\bm{k}}-E_{\beta\bm{k}})[\bm{a}_{\text{lat}}(\bm{k})]_{\beta\alpha}f_{\alpha}\nonumber \\
 & +\text{Re}\sum_{\bm{k}\alpha ll^\prime pp^\prime}\sum_{\beta\neq\alpha}(\phi_{\alpha\bm{k}l}^{\ast}\bm{P}_{lp}(\bm{k})\phi_{\beta\bm{k}p})\times(\phi_{\beta\bm{k}p^\prime}^{\ast}\bm{v}_{p^\prime l^\prime}(\bm{k})\phi_{\alpha\bm{k}l^\prime})f_{\alpha}\nonumber \\
 & +\frac{i}{2e\hbar}\sum_{\bm{k}\alpha ll^\prime p}\phi_{\alpha\bm{k}l}^{\ast}(\bm{P}_{lp}(\bm{k})\times[\bm{P}(\bm{k}),H(\bm{k})]_{pl^\prime})\phi_{\alpha\bm{k}l^\prime}f_{\alpha},
\end{align}
where we have introduced the velocity operator for the lattice Hamiltonian
as 
\begin{align}
\bm{v}_{ll^\prime}(\bm{k}) & =\frac{1}{\hbar}\frac{\partial H_{ll^\prime}}{\partial\bm{k}},
\end{align}
whose matrix element in the band basis is given by 
\begin{align}
\sum_{ll^\prime}\phi_{\alpha\bm{k}l}^{\ast}\bm{v}_{ll^\prime}(\bm{k})\phi_{\beta\bm{k}l^\prime} & =\frac{1}{\hbar}\frac{\partial E_{\alpha\bm{k}}}{\partial\bm{k}}\delta_{\alpha\beta}-\frac{i}{\hbar}(E_{\beta\bm{k}}-E_{\alpha\bm{k}})[\bm{a}_{\text{lat}}]_{\alpha\beta}.
\end{align}
We have also introduced a short-hand notation $[A,B]_{ll^\prime}=\sum_{p}(A_{lp}B_{pl^\prime}-B_{lp}A_{pl^\prime})$.
The same relation for the electric dipole can also be used for the
second term $\bm{M}^{(2)}$, which results in
\begin{align}
\bm{M}^{(2)} & =-\frac{ek_{B}T}{\hbar}\sum_{\bm{k}\alpha}\frac{\partial}{\partial\bm{k}}\times[\bm{a}_{\text{lat}}(\bm{k})]_{\alpha\alpha}\ln(1-f_{\alpha})\nonumber \\
 & +\sum_{\bm{k}\alpha ll^\prime pp^\prime}(\phi_{\alpha\bm{k}l}^{\ast}\bm{P}_{lp}(\bm{k})\phi_{\alpha\bm{k}p})\times(\phi_{\alpha\bm{k}p^\prime}^{\ast}\bm{v}_{p^\prime l^\prime}(\bm{k})\phi_{\alpha\bm{k}l^\prime})f_{\alpha}.
\end{align}
Here, we have performed integration by part for the Wannier correction
term, and used the relation
\begin{align}
\frac{\partial}{\partial\bm{k}}\ln(1-f_\alpha) & =\frac{1}{k_{B}T}\frac{\partial E_{\alpha\bm{k}}}{\partial\bm{k}}f_{\alpha}.
\end{align}
Note that the Wannier correction for $\bm{M}^{(2)}$ has the same
form as that for $\bm{M}^{(1)}$. By combining these terms, we arrive
at 
\begin{align}
\bm{M}_{\text{orb}} &=\bm{M}_{\text{lat}}+\text{Re}\sum_{\bm{k}ll^\prime p}\left\langle c_{\bm{k}l}^{\dagger}\bm{P}_{lp}(\bm{k})\times\bm{v}_{pl^\prime}(\bm{k})c_{\bm{k}l^\prime}\right\rangle\nonumber\\
&+\frac{i}{2e\hbar}\sum_{\bm{k}ll^\prime p}\left\langle c_{\bm{k}l}^{\dagger}\bm{P}_{lp}(\bm{k})\times[\bm{P}(\bm{k}),H(\bm{k})]_{pl^\prime}c_{\bm{k}l^\prime}\right\rangle .
\end{align}
We note that the summation on the orbital $l$ is not restricted to
the entangled bands. To restrict the range of the orbital summation,
let us rewrite the expression with the matrix element of the orbital
angular momentum Eq.~(\ref{eq:angular-momentum-wannier}). By inserting the resolution of
the identity $\sum_{il}w_{l}(\bm{r}-\bm{R}_{i})w_{l}^{\ast}(\bm{r}^{\prime}-\bm{R}_{i})=\delta(\bm{r}-\bm{r}^{\prime})$,
we obtain 
\begin{align}
\bm{L}_{ll^\prime} & =-\frac{m_{e}}{2e}\sum_{p}(\bm{P}_{l^\prime p}\times\bm{V}_{pl^\prime}-\bm{V}_{lp}\times\bm{P}_{pl^\prime})
\end{align}
where $\bm{V}_{ll^\prime}$ is the matrix element of the velocity operator
in the continuous space, which satisfies
\begin{align}
\bm{V}_{ll^\prime}(\bm{k}) & =\sum_{i}\int d^{3}\bm{r}w_{l}^{\ast}(\bm{r})\frac{-i\hbar}{m_{e}}\bm{\nabla}w_{l^\prime}(\bm{r}+\bm{R}_{i})e^{-i\bm{k}\cdot\bm{R}_{i}}\\
 & =\bm{v}_{ll^\prime}(\bm{k})+\frac{i}{e\hbar}[\bm{P}(\bm{k}),H(\bm{k})]_{ll^\prime}.
\end{align}
Namely, we finally obtain the expression for the orbital magnetization
as
\begin{align}
\bm{M}_{\text{orb}} & =\bm{M}_{\text{lat}}-\frac{e}{2m_{e}}\sum_{\bm{k}ll^\prime}\langle c_{\bm{k}l}^{\dagger}\bm{L}_{ll^\prime}c_{\bm{k}l^\prime}\rangle\nonumber\\
&+\frac{1}{2}\text{Re}\sum_{\bm{k}ll^\prime p}\left\langle c_{\bm{k}l}^{\dagger}\bm{P}_{lp}(\bm{k})\times\bm{v}_{pl^\prime}(\bm{k})c_{\bm{k}l^\prime}\right\rangle ,
\end{align}
which is nothing but Eq.~(\ref{eq:orbitalmag-decomp}). Here, we note that the orbital
sum $p$ can be restricted to the entangled bands, since the velocity
operator $\bm{v}_{ll^\prime}$ only has nonzero matrix elements within the
entangled bands.

\section{Explicit form of the Floquet Hamiltonian}\label{sec:appendix-FloquetHamiltonian}

In this appendix, we explicitly present the matrix representation of the Floquet Hamiltonian for the $s$-$p$ tight-binding model.
By performing Fourier transformation with the Jacobi-Anger identity 
\begin{align}
    e^{iz\cos \theta} = \sum_{n=-\infty}^{\infty}i^n \mathcal{J}_n(z)e^{in\theta},
\end{align}
where $\mathcal{J}_n$ is the $n$th Bessel function,
we obtain the explicit form of the matrix element of the Floquet Hamiltonian
$H_{n,\bm{k},\sigma}=T^{-1}\int_0^T dt H_{\bm{k},\sigma}(t)e^{in\Omega t}$  as
\begin{align}
    H_{n,\bm{k},\sigma} &= 
    \begin{pmatrix}
        E^s_n & V^x_n & V^y_n \\
        (V^x_{-n})^{*} & \Delta \delta_{n,0} + E^x_n & -\frac{i}{2}\lambda\sigma\delta_{n,0} \\
        (V^y_{-n})^{*} & \frac{i}{2}\lambda\sigma\delta_{n,0} & \Delta \delta_{n,0} +E^y_n
    \end{pmatrix}
    ,
\end{align}
with
\begin{subequations}
\begin{align}
    E^s_n =& -t_s \left[
        e^{ik_x} + (-1)^n e^{-ik_x} + i^n e^{ik_y} + (-i)^n e^{-ik_y}
        \right]
        \mathcal{J}_n(A), \\
    E^{x/y}_n =& -t_{a/b} \left[
        e^{ik_x} + (-1)^n e^{-ik_x}
    \right]
    \mathcal{J}_n(A) \n
    &-t_{b/a} \left[
        i^n e^{ik_y} + (-i)^n e^{-ik_y}
    \right]
    \mathcal{J}_n(A), \\
    V^x_n =& -t_{sp} \left[
        e^{ik_x} - (-1)^n e^{-ik_x}
    \right]
    \mathcal{J}_n(A) - \frac{PE}{2}(\delta_{n,1} + \delta_{n,-1}), \\
    V^y_n =&  -t_{sp} \left[
        i^n e^{ik_y} - (-i)^n e^{-ik_y}
    \right]
    \mathcal{J}_n(A) - \frac{PE}{2}i(\delta_{n,1} - \delta_{n,-1}).
\end{align}
\end{subequations}

\section{Effect of magnetic field}\label{sec:appendix-magnetic-field}

In this Appendix, we briefly discuss the effect of a magnetic field on the IFE. 
The generation of spin magnetization due to the Zeeman coupling has been discussed in the previous literature \cite{Takayoshi2014,Takayoshi2014-2}. Here, we briefly review the essential argument. 
The Hamiltonian describing the coupling between the oscillating magnetic field and spin is given by
\begin{align}
    H^{B}(t) = \mu_B \bm{B}(t) \cdot \bm{S},
\end{align}
where $\bm{B}(t) = B(\cos \Omega t, \sin \Omega t)$ is the magnetic field. 
The effective Hamiltonian obtained via high-frequency expansion is
\begin{align}
    H^B_{\text{eff}} = \sum_{n > 0}\frac{[H^B_{-n},H^B_n]}{n\hbar \Omega} 
    = -\frac{\mu_B^2 B^2S^z}{2\hbar \Omega},
\end{align}
where we defined $H^B_n = T^{-1}\int dt H^B(t)e^{in\Omega t}$.
This indicates that spin magnetization is induced along the $z$-axis.
Substituting $B = E/c$, where $E$ is the strength of the electric field and $c$ is the speed of light, the effective magnetic field is given by
\begin{align}
    B_{\text{eff}} = -\frac{\mu_B E^2}{2\hbar\Omega c^2}.
\end{align}
Using $E = 1$ MV/cm and $\hbar\Omega = 6$ eV, the effective magnetic field is estimated to be $B_{\text{eff}} = 5\times 10^{-7}$ T. This value is significantly smaller compared to the estimate of $B_{\text{eff}} = 30$ T in Sec.~\ref{sec : Discussion}, suggesting that the effect of the magnetic field is negligible compared to that of the electric field in this setup.

\section{Derivation of the distribution for Floquet states}\label{sec:appendix-distribution}

In this appendix, we derive the distribution function for Floquet states based on the Floquet Keldysh formalism~\cite{Aoki2014,Seetharam2015,Morimoto2016}.
The retarded, advanced and lesser components of the Green's function are defined as
\begin{align}
    G^{R}_{\bm{k},\alpha\beta}(t,t') &= 
    -i\theta(t-t')\braket{\{c_{\bm{k},\alpha}(t),c_{\bm{k},\beta}^{\dagger}(t')\}}, \\
    G^{A}_{\bm{k},\alpha\beta}(t,t') &=
    i\theta(t'-t)\braket{\{c_{\bm{k},\alpha}(t),c_{\bm{k},\beta}^{\dagger}(t')\}}, \\
    G^{<}_{\bm{k},\alpha\beta}(t,t') &=
    i\braket{c_{\bm{k},\beta}^{\dagger}(t')c_{\bm{k},\alpha}(t)},
\end{align}
where $c_{\bm{k},\alpha}$ $(c^{\dagger}_{\bm{k},\alpha})$ is the annihilation (creation) operator for the system's fermions and $\alpha$ denotes the internal degree of freedom.

We consider a model where the system is coupled to a fermionic reservoir at each site. The Hamiltonian reads
\begin{align}
    H_{\text{tot}}(t) &= H_{\text{sys}}(t) + H_{\text{mix}} + H_{\text{bath}}, \\ 
    H_{\text{sys}}(t) &= \sum_{\bm{k},\alpha\beta}
    \left[ H_{\bm{k}}(t) \right]_{\alpha\beta}
    c_{\bm{k},\alpha}^{\dagger}c_{\bm{k},\beta}, \\
    H_{\text{mix}} &= \sum_{\bm{k},\alpha,p}(V_{p}c_{\bm{k},\alpha}^{\dagger}b_{\bm{k},p} + \text{h.c.}), \\
    H_{\text{bath}} &= \sum_{\bm{k},p}\omega_p b_{\bm{k},p}^{\dagger}b_{\bm{k},p},
\end{align}
where $\omega_p$ is the level energy in the reservoir, $b_{\bm{k},p}$ $(b^{\dagger}_{\bm{k},p})$ is the annihilation (creation) operator for the fermions in the reservoir and $V_p$ is the coupling strength between the system and the reservoir.
The retarded and advanced component of the self energy is written using Green's function for the reservoir as
\begin{align}
    \Sigma^{R}_{\bm{k}}(t,t') &= -i\theta(t-t')\sum_{p}|V_p|^2\braket{\{b_{\bm{k},p}(t),b_{\bm{k},p}^{\dagger}(t')\}}_{\text{bath}} \n
    &= -i\theta(t-t')\sum_{p}|V_p|^2 e^{-i\omega_p (t-t')}, \\
    \Sigma^{A}_{\bm{k}}(t,t') &= i\theta(t'-t)\sum_{p}|V_p|^2\braket{\{b_{\bm{k},p}^{\dagger}(t'),b_{\bm{k},p}(t)\}}_{\text{bath}} \n
    &= i\theta(t'-t)\sum_{p}|V_p|^2 e^{-i\omega_p (t-t')}.
\end{align}
Hereafter, we omit the subscript $\bm{k}$ for simplicity. 

Let us introduce the Floquet representation of the Green's function. We assume that the Green's function is periodic in time, i.e. $G(t+T,t'+T) = G(t,t')$. The Floquet representation is written as
\begin{align}
    G_{mn}(\omega) &= \int^{T}_{0}\frac{dt_a}{T}\int^{\infty}_{-\infty}dt_{r}
    e^{i(m-n)\Omega t_a}
    e^{i(\omega + \frac{m+n}{2}\Omega)t_r},
\end{align}
where we define $t_a = (t+t')/2, t_r = t-t'$. Then, we have
\begin{align}
    \Sigma_{mn}^{R/A}(\omega) &= \sum_{p} \frac{|V_p|^2 \delta_{nm}}{\omega + n\Omega - \omega_p \pm i\delta}.
\end{align}
Using the relation $\frac{1}{\omega + i\delta} = \mathcal{P}\frac{1}{\omega} - i\pi\delta(\omega)$, we obtain the imaginary part of the self energy as
\begin{align}
    \text{Im}\Sigma_{mn}^{R/A}(\omega) &= \mp \sum_{p} \pi|V_p|^2\delta(\omega + n\Omega - \omega_p)\delta_{mn},
\end{align}
which corresponds to the spectrum of the reservoir.
We adopt the broad-band condition
\begin{align}
    \sum_p \pi |V_p|^2 \delta(\omega - \omega_p) = \Gamma,
\end{align}
and neglect $\omega$ dependence of $\text{Im}\Sigma_{mn}^{R/A}(\omega)$. The real part $\text{Re}\Sigma_{mn}^{R/A}(\omega)$
can be absorbed in the renormalization of the energy levels. Thus, the self energy is given by
\begin{align}
    \Sigma_{mn}^{R/A}(\omega) &= \mp i\Gamma \delta_{mn}.
\end{align}
We consider the situation where the reservoir is so large that it is in the equilibrium with the temperature $T = \beta^{-1}$. Then, the following fluctuation dissipation theorem holds~\cite{Aoki2014}:
\begin{align}
    \Sigma^{K}_{mn}(\omega) &= \tanh\frac{\beta(\omega + m\Omega)}{2}
    \left(
        \Sigma^{R}_{mn}(\omega) - \Sigma^{A}_{mn}(\omega)
    \right) 
    \delta_{mn}.
\end{align}
Combined with the relation
$\Sigma^{<} = (\Sigma^{A}-\Sigma^{R}+\Sigma^{K})/2$, 
the lesser component of the self energy is 
\begin{align}
    \Sigma^{<}_{mn}(\omega)
    &= 2i\Gamma f(\omega + n\Omega)\delta_{mn}.
    \label{eq: Lesser self energy}
\end{align}

Let us calculate the retarded and advanced component of the Green's function with the self energy derived above.
The inverse of the bare Green's function is given by
\begin{align}
    (G^{R/A}_{0}(\omega))^{-1}_{mn} &= 
    (\omega + m\Omega \pm i \delta)\delta_{mn} - H_{mn},
\end{align}
where we define 
$H_{mn} = T^{-1}\int_0^T dt e^{i(m-n)\Omega t}H(t)$.
By using the Dyson equation, 
\begin{align}
    (G^{R/A}(\omega))^{-1}_{mn}
    = 
    (G^{R/A}_{0}(\omega))^{-1}_{mn}
    -
   \Sigma_{mn}^{R/A}(\omega),
\end{align}
the inverse of the Green's function is obtained as
\begin{align}
    (G^{R/A}(\omega))^{-1}_{mn} &= 
    (\omega + m\Omega \pm i \Gamma)\delta_{mn} - H_{mn}.
\end{align}
Using the basis $\ket{u_{m,\alpha}}$ that diagonalizes the Floquet Hamiltonian, which is explicitly written as
Eq.~(\ref{eq:sambe}),
Green's function is expressed as
\begin{align}
    G_{mn}^{R/A}(\omega) &= \sum_{\alpha} \frac{\ket{u_{m,\alpha}}\bra{u_{n,\alpha}}}{\omega  - \epsilon_{
    \alpha} \pm i\Gamma}. \label{eq: RA Green}
\end{align}

Now we can calculate the lesser component of the Green's function given by
\begin{align}
    G^{<}(t,t') &= \int d\tau \int d\tau' G^{R}(t,\tau)\Sigma^{<}(\tau,\tau')G^{A}(\tau',t').
\end{align}
Using Eq.~(\ref{eq: Lesser self energy}) and Eq.~(\ref{eq: RA Green}), $G_{mn}^{<}(\omega)$ is expressed as
\begin{align}
    G_{mn}^{<}(\omega)
    &= 
    \sum_{l,l'}
    G^{R}_{ml}(\omega)
    \Sigma_{ll'}^{<}(\omega)
    G^{A}_{l'n}(\omega) \n
    &= 
    \sum_{l,\alpha,\beta}
    \frac{2i\Gamma f(\omega + l\Omega)\braket{u_{l,\alpha}|u_{l,\beta}}}{(\omega - \epsilon_{\alpha} + i\Gamma)(\omega  - \epsilon_{\beta} - i\Gamma)} \ket{u_{m,\alpha}}\bra{u_{n,\beta}}.
\end{align}
With this expression, the time average of the expectation value of the observable $\mathcal{O}(t)$ is calculated as
\begin{align}
    \overline{\braket{\mathcal{O}(t)}}
    &= 
    \frac{1}{T} \int_0^T dt 
    \sum_{\bm{k}}
    \text{Tr}
    \left[
        \mathcal{O}(t) (-i)G^{<}(t,t)
    \right] 
    \n
    &= 
    \int \frac{d\omega}{2\pi}
    \sum_{\bm{k}}
    \text{Tr}
    \left[
        \sum_{m,n}
        \mathcal{O}_{nm}
        (-i)
        G_{mn}^{<}(\omega)
    \right]
    \n
    &=
    \sum_{\bm{k},l,\alpha,\beta} 
    \int \frac{d\omega}{2\pi}
    \frac{2\Gamma f(\omega + l\Omega)\braket{u_{l,\alpha}|u_{l,\beta}}}{(\omega - \epsilon_{\alpha} + i\Gamma)(\omega  - \epsilon_{\beta} - i\Gamma)} \n
    &\quad \quad \quad \quad \times
        \sum_{m,n}\bra{u_{n,\beta}}\mathcal{O}_{nm} \ket{u_{m,\alpha}}
    \n
    &\xrightarrow{\Gamma\to 0}
    \sum_{\bm{k},l,\alpha}
    f(\epsilon_{\alpha}+l\Omega)\braket{u_{l,\alpha}|u_{l,\alpha}} 
        \sum_{m,n}\bra{u_{n,\alpha}}\mathcal{O}_{nm} \ket{u_{m,\alpha}},
\end{align}
which reproduces Eqs.~(\ref{eq: Floquet distribution}) and (\ref{eq: Floquet expectation value}).
Here we have introduced $\mathcal{O}_{mn} = T^{-1}\int_0^T dt \mathcal{O}(t)e^{i(m-n)\Omega t}$ and taken the limit $\Gamma\to 0$ by using the relation
\begin{align}
    \int \frac{d\omega}{2\pi}\frac{2\Gamma}{(\omega - \epsilon_{\alpha}+ i\Gamma)(\omega - \epsilon_{\beta}- i\Gamma)} &\xrightarrow{\Gamma\to 0} \delta_{\alpha\beta},
\end{align}
under an assumption that the Floquet states are nondegenerate.

\section{Schrieffer-Wolff transformation}\label{sec:appendix-SW}

In this appendix, we give a detailed calculation of the Schrieffer-Wolff transformation~\cite{Bukov2016,Kitamura2017,Schrieffer1966}. 

The block-matrix form of the Bloch Hamiltonian and its Fourier component is given by
\begin{align}
    H(t) &= 
    \begin{pmatrix}
        E^s(t) & V(t) \\
        V^{\dagger}(t) & E^p(t)
    \end{pmatrix}
    ,\,
    H_{n} = \int_0^T \frac{dt}{T}H(t)e^{in\Omega t}=
    \begin{pmatrix}
        E^s_n & V_n \\
        V^{\dagger}_{-n} & E^p_n
    \end{pmatrix}
\end{align}
with
\begin{align}
    V(t)
    = 
    \begin{pmatrix}
        V^x(t) & V^y(t)
    \end{pmatrix},\,
    E^p(t)
    = 
    \begin{pmatrix}
        \Delta + E^x(t) & -\frac{i}{2}\lambda\sigma \\
        \frac{i}{2}\lambda\sigma & \Delta + E^y(t)
    \end{pmatrix}.
\end{align}
Here, we note that $E^s(t),E^s_n$ are $1\times1$ matrices, while $E^p(t),E^p_n$ are $2\times 2$ matrices. We regard $V(t),V_n$ as a perturbation to the Hamiltonian.

Given a matrix $S(t)$ satisfying $S(t+T)=S(t)$, the Schrieffer-Wolff transformation for the time-periodic system is given by
\begin{align}
    H_{\text{SW}}(t) &= e^{iS(t)}\left(H(t) - i\partial_t\right)e^{-iS(t)} \n
    &= H(t) 
    + [iS(t),H(t)-i\partial_t] \n
    &+ \frac{1}{2}[iS(t),[iS(t),H(t)-i\partial_t]] + \cdots 
\end{align}
By decomposing $S(t)$ into $S(t) = S^{(1)}(t) + S^{(2)}(t) + S^{(3)}(t) + \cdots$, where $S^{(n)}(t)$ is the $n$th order perturbation, we can write the effective Hamiltonian by order by order:
\begin{align}
    H_{\text{SW}}^{(1)}(t) &= H^{(1)}(t) + [iS^{(1)}(t),H^{(0)}(t)-i\partial_t], \label{eq: perturbation1} \\ 
    H_{\text{SW}}^{(2)}(t) &= [iS^{(1)}(t),H^{(1)}(t)-i\partial_t] + [iS^{(2)}(t),H^{(0)}(t)-i\partial_t] \n
    &+ \frac{1}{2}[iS^{(1)}(t),[iS^{(1)}(t),H^{(0)}(t)-i\partial_t]], \label{eq: perturbation2}
\end{align}
where $H^{(0)}(t)$ and $H^{(1)}(t)$ are the unperturbed and the perturbed Hamiltonian, respectively.

Upon transformation into Fourier space, Eq.~(\ref{eq: perturbation1}) becomes
\begin{align}
    H_n^{(1)} + i\sum_m [S^{(1)}_m,H_{n-m}^{(0)}] + in\Omega S^{(1)}_n = 0
    \label{eq: Perturbation H1}
\end{align}
By introducing the matrix form of $S^{(1)}(t)$,
\begin{align}
    S^{(1)}(t) &=
    \begin{pmatrix}
        0 & T(t) \\
        T^{\dagger}(t) & 0
    \end{pmatrix},
\end{align}
with $T(t)$ satisfying $T(t+T) = T(t)$,
Eq.~(\ref{eq: Perturbation H1}) can be written as
\begin{align}
    V_n + i\sum_m(T_m E^{p}_{n-m} - E^{s}_{n-m}T_{m}) + in\Omega T_n = 0.
\end{align}
For simplicity, we drop $E_{n}^{p},E_{n}^{s}\ (n \neq 0)$ since they are higher order in terms of the electric field. We can determine $T_n$ as
\begin{align}
    &V_n + iT_n\left[E^{p}_{0}-(E^{s}_{0}-n\Omega)I_{2\times 2}\right] = 0, \\
    \ & T_n = iV_n\left[E^{p}_{0}-(E^{s}_{0}-n\Omega)I_{2\times 2}\right]^{-1}. \label{eq: Tn}
\end{align}
With this $T_n$, we have
\begin{align}
    H^{(1)}(t) + [iS^{(1)}(t),H^{(0)}(t)-i\partial_t] = 0.
    \label{eq: perturbation1_2}
\end{align}
Substituting Eq.~(\ref{eq: perturbation1_2}) into Eq.~(\ref{eq: perturbation2}) gives
\begin{align}
    H_{\text{SW}}^{(2)}(t) &= \frac{1}{2}[iS^{(1)}(t),H^{(1)}(t)-i\partial_t] + [iS^{(2)}(t),H^{(0)}(t)-i\partial_t].
\end{align}
$S^{(2)}(t)$ is determined so as to eliminate the offdiagonal term of $H_{\text{SW}}^{(2)}(t)$, thus we have
\begin{align}
    \mathcal{P}H_{\text{SW}}^{(2)}(t)\mathcal{P} &= \frac{1}{2}\mathcal{P}[iS^{(1)}(t),H^{(1)}(t)-i\partial_t]\mathcal{P}, \\
    \mathcal{P}[H_{\text{SW}}^{(2)}]_n \mathcal{P}
    &=
    \frac{1}{2}
    \mathcal{P}
    \left[
    i\sum_m [S^{(1)}_m, H_{n-m}^{(1)}] + in\Omega S^{(1)}_n 
    \right]
    \mathcal{P} \n
    &=
    \frac{1}{2}
        i\sum_m (T_m V^{\dagger}_{m-n} - V_{n-m}T^{\dagger}_{-m}),
\end{align}
where $\mathcal{P}=\text{diag}(1,0,0)$ is the projector onto the $s$ orbital.
Substituting $n=0$ and Eq.~(\ref{eq: Tn}) into the above equation, we obtain Eq.~(\ref{eq: analytical Heff}).

The effective angular momentum operator can be obtained by the same unitary transformation
\begin{align}
    L_{\text{SW}}^z(t) &= e^{iS(t)}L^z e^{-iS(t)} \n
    &= L^z + [iS(t),L^z] + \frac{1}{2}[iS(t),[iS(t),L^z]] + \cdots.
\end{align}
Under the second order perturbation, we have
\begin{align}
    \mathcal{P}L_{\text{SW}}^{(2)z}(t)\mathcal{P} &= 
    \frac{1}{2}\mathcal{P}[iS^{(1)}(t),[iS^{(1)}(t),L^z]]\mathcal{P}, \\
    \mathcal{P}\left[L_{\text{SW}}^{(2)z}\right]_n\mathcal{P}
    &=  
    \frac{1}{2}\sum_m (T_{n-m}L^z T^{\dagger}_{-m} + T_m L^z T^{\dagger}_{m-n}).
\end{align}
Substituting $n=0$ and Eq.~(\ref{eq: Tn}) into the above equation, we obtain Eq.~(\ref{eq: Lzeff}).

For the orbital magnetic moment from Wannier polarization contribution, we define the matrix element and its Fourier component as
\begin{align}
    M^z_{P}(t) &= \frac{1}{4}
    \left(
        P_x \frac{\partial H(t)}{\partial k_y} - P_y \frac{\partial H(t)}{\partial k_x}
    \right)+\text{h.c.}
    =
    \begin{pmatrix}
        0 & W(t) \\
        W^{\dagger}(t) & F^p(t)
    \end{pmatrix},
    \n
    \left[M^z_P\right]_n &= \int_{0}^{T}\frac{dt}{T}M^z_P(t)e^{in\Omega t} = 
    \begin{pmatrix}
        0 & W_n \\
        W^{\dagger}_n & F^p_n
    \end{pmatrix}.
\end{align}
In the same way as the orbital angular momentum, we can obtain the effective operator for Wannier polarization contribution by the unitary transformation
\begin{align}
    M_{P,\text{SW}}^z(t) &= e^{iS(t)}M^z_P(t) e^{-iS(t)} \n
    &= M^z_P(t) + [iS(t),M^z_P(t)] \n
    & + \frac{1}{2}[iS(t),[iS(t),M^z_P(t)]] + \cdots.
\end{align}
Under the first order perturbation, we obtain
\begin{align}
    \mathcal{P}M_{P,\text{SW}}^{(1)z}(t)\mathcal{P}
    &= \mathcal{P}\left[ iS^{(1)}(t), M_{P,\text{SW}}^{(1)z}(t)\right] \\
    \mathcal{P}\left[ M_{P,\text{SW}}^{(1)z} \right]_n \mathcal{P} &=i \sum_{m}(T_m W_{m-n}^{\dagger} - W_{n-m}T_{-m}^{\dagger})
\end{align}
Substituting $n=0$ and Eq.~(\ref{eq: Tn}) into the above equation, we obtain Eq.~(\ref{eq: MzPeff}).


\bibliography{ref.bib}

\end{document}